# High-Entropy A-Site Disorder Enhances Oxygen Evolution Reaction Activity and Stability in Nickelate Thin Films

*Min-Ju Choi, Le Wang*, Krishna Prasad Koirala, Hyoju Park, Mark E. Bowden, Elizabeth Helen Krenkel, Aiping Chen, Jueli Shi, Conan Weiland, Iradwikanari Waluyo, Adrian Hunt, Peter V. Sushko, and Yingge Du**

Dr. M. Choi, Dr. L. Wang, Dr. K. P. Koirala, Dr. H. Park, Dr. M. E. Bowden, Dr. J. Shi, Dr. P. V. Sushko, Dr. Y. Du
Integrated Discovery Sciences Directorate, Pacific Northwest National Laboratory, Richland, Washington 99354, United States
E-mail: le.wang@pnnl.gov, yingge.du@pnnl.gov
Dr. K.P. Koirala
Materials Department, University of California Santa Barbara, Santa Barbara, California 93106, United States
Dr. E. H. Krenkel, Dr. A. Chen
Center for Integrated Nanotechnologies, Los Alamos National Laboratory, Los Alamos, New Mexico 87545, United States
Dr. C. Weiland
Material Measurement Laboratory, National Institute of Standards and Technology, Gaithersburg, Maryland 20899, United States
Dr. I. Waluyo, Dr. A. Hunt
National Synchrotron Light Source II, Brookhaven National Laboratory, Upton, New York 11973, United States

**Abstract:** Rare-earth nickelates are among the most active oxide electrocatalysts for the oxygen evolution reaction (OER), owing to their tunable Ni–O covalency, near-unity $e_g$ occupancy, and sensitivity to A-site chemistry, oxygen stoichiometry, and epitaxial strain. Here, we use epitaxial A-site high-entropy $(La_{0.2}Pr_{0.2}Nd_{0.2}Sm_{0.2}Eu_{0.2})NiO_3$ thin films to reveal how epitaxial strain influences defect formation, Ni valence, electronic structure, and OER activity in a chemically disordered rare-earth (RE) sublattice. Tensile strain promotes Ruddlesden–Popper fault formation, interrupts the corner-sharing $NiO_6$ network, and lowers the average Ni oxidation state. Although the occupied density of states near the Fermi level remains largely unchanged, O K-edge X-ray absorption spectroscopy reveals increased low-energy unoccupied states under tensile strain. First-principles calculations suggest that tensile strain shifts RE 4f-derived states toward the Fermi level, enhancing RE 4f/O 2p/Ni 3d hybridization. These coupled structural and electronic changes lead to earlier OER onset, stronger pre-catalytic Ni redox features, and higher activity than compressively strained films and representative single-A-site nickelates. The high-entropy nickelate also shows enhanced resistance to OER-induced structural degradation compared with $LaNiO_3$. These results suggest that A-site configurational disorder and epitaxial strain can jointly tune defect formation and rare-earth-derived electronic states, providing a route for improving perovskite oxide electrocatalysts.

## 1. Introduction

A fundamental understanding of the factors governing oxygen evolution reaction (OER) activity and stability is essential for the predictive design of oxide electrocatalysts that are both highly active and durable.[1-5] Among oxide electrocatalyst candidates, $ABO_3$-type perovskites have attracted sustained attention because their electronic structures, and consequently their catalytic properties, can be systematically tuned through cation substitution, oxygen stoichiometry control, epitaxial strain, and surface/interface engineering.[6-9] Within this family, rare-earth perovskite nickelates ($RENiO_3$, where RE denotes a rare-earth element) are particularly promising OER electrocatalysts.[10, 11] For example, $LaNiO_3$ (LNO) is among the most active perovskite OER electrocatalysts, a property commonly associated with efficient charge transfer arising from strong Ni 3d-O 2p hybridization, $e_g$ occupancy close to unity, and relatively weak OH* adsorption compared with many other first-row transition metal oxides.

A-site chemical substitution provides a useful platform for elucidating how structural and electronic factors influence OER activity in nickelates, as it modifies lattice geometry and electronic structure without directly altering the catalytically active Ni sites.[12, 13] Prior studies on $La_{1-x}RE_xNiO_3$ found that partial replacement of La with smaller rare-earth elements such as Nd or Sm incorporates more oxygen vacancies ($V_O^{\cdot\cdot}$, where two dots ·· indicate two positive charges relative to the neutral lattice site) under the same synthesis conditions, reducing the Ni valence state from $Ni^{3+}$ to $Ni^{2+}$ and increasing the average $e_g$ orbital filling, favorable for OER.[13] Extending this concept, high-entropy A-site disorder introduces an additional dimension to this design space.[14-18] In high-entropy perovskite nickelates, incorporating five or more rare-earth elements into the A-site sublattice introduces substantial local variations in ionic size, bonding environment, lattice distortion, and rare-earth electronic structure while preserving the catalytically active Ni-O framework.[18, 19] Compared with conventional single-A-site or lightly alloyed $RENiO_3$, such configurational disorder may stabilize otherwise inaccessible structural states, alter oxygen-vacancy energetics, broaden local Ni–O–Ni bond distributions, and introduce multiple rare-earth 4f-derived electronic states near the Fermi level ($E_F$). Prior work has shown that epitaxial $(La_{0.2}Pr_{0.2}Nd_{0.2}Sm_{0.2}Eu_{0.2})NiO_3$ (hereafter denoted $(5A)NiO_3$) nickelate films can be synthesized as single-phase perovskite-like films with predominantly $Ni^{3+}$ character despite extreme A-site disorder.[18, 20, 21] However, how this high-entropy A-site disorder affects defect formation (e.g., $V_O^{\cdot\cdot}$ and Ruddlesden–Popper (RP) faults), the electronic structure, and OER functionality remains largely unexplored.

Varying epitaxial strain offers a complementary approach to probe the influence of these disorder-coupled effects on electrocatalytic behavior. In conventional single-A-site $RENiO_3$, like LNO, epitaxial strain has been reported to tune the electronic structure, including orbital polarization and Ni-O hybridization, thereby influencing OER activity when $V_O^{\cdot\cdot}$ are minimized.[22, 23] Moving from LNO to $NdNiO_3$ (NNO), strain-dependent OER behavior becomes further complicated by the strong coupling between epitaxial lattice distortion and $V_O^{\cdot\cdot}$ formation, especially under tensile strain.[24] Beyond homogeneous lattice distortion, tensile strain in perovskite nickelates can also promote the formation of RP faults, which introduce locally symmetry-broken and chemically distinct motifs that may serve as active catalytic sites.[25] These findings suggest that strain does not simply tune elastic lattice parameters; it can also strongly affect the defect formation, electronic structure, and surface catalytic behavior. However, whether these strain-mediated mechanisms persist, compete, or evolve in A-site high-entropy nickelates with chemically disordered rare-earth sublattices remains unclear.

Moreover, recent studies on fluorite $CeO_2$ further highlight that strain can influence electrocatalytic behavior through electronic states beyond conventional transition-metal d bands.[26] In tensile-strained $CeO_2$ thin films, the Ce 4f band can be shifted toward the $E_F$, altering metal–oxygen hybridization and redox flexibility. This concept is particularly relevant for $(5A)NiO_3$, where multiple rare-earth A-site cations possess partially filled or low-lying 4f states. If epitaxial strain can modulate the energetic alignment of these rare-earth 4f states, then A-site disorder may do more than impose structural distortion: it may actively contribute to the unoccupied electronic structure involved in charge transfer during OER. Yet the role of rare-earth 4f states, their coupling to Ni 3d/O 2p hybridized bands, and their strain-dependence in high-entropy nickelate electrocatalysts remain to be revealed.

Here, we use epitaxial (5A)$NiO_3$ as a model system to elucidate how high-entropy A-site disorder, epitaxial strain, defect formation, and electronic-structure modulation collectively govern OER activity. We grow (5A)$NiO_3$ thin films on three different (001)-oriented substrates, including $LaAlO_3$ (LAO) (compressive strain), $(LaAlO_3)_{0.3}(Sr_2AlTaO_6)_{0.7}$ (LSAT) (moderate tensile strain), and $SrTiO_3$ (STO) (tensile strain). Structural characterization and first-principles calculations show that tensile strain promotes RP-fault formation by lowering the energetic penalty. These RP-type faults introduce rock-salt-like A–O layers that disrupt the corner-sharing $NiO_6$ network and favor a more $Ni^{2+}$-like character, consistent with the systematic decrease in average Ni oxidation state observed by Ni L-edge X-ray absorption spectroscopy (XAS). Although the occupied density of states (DOS) near the $E_F$ remains largely unchanged, O K-edge XAS and theoretical calculations reveal that tensile strain strongly modifies the unoccupied DOS by drawing rare-earth 4f-derived states closer to $E_F$, thereby increasing the low-energy unoccupied states and enhancing RE 4f/O 2p/Ni 3d hybridization. Electrochemical measurements show that tensile-strained (5A)$NiO_3$ films exhibit earlier OER onset, stronger pre-catalytic Ni redox features, and substantially higher activity than the compressively strained film and most reported single-A-site nickelate thin films. During stability testing, the (5A)$NiO_3$ film exhibits a more stable electrochemical response and better preservation of its crystalline framework than $LaNiO_3$, indicating enhanced resistance to OER-induced structural degradation. These results establish A-site high-entropy disorder as an effective strategy for coupling strain-tunable defect formation, rare-earth-derived electronic states, and improved structural robustness in perovskite oxide electrocatalysts.

## 2. Results and Discussion

(5A)$NiO_3$ thin films (11 nm to 20 nm) were grown on (001)-oriented LAO, LSAT, and STO substrates (a = 3.79 Å, 3.868 Å, and 3.905 Å, respectively) using pulsed laser deposition (PLD) (See Methods for details). Bulk (5A)$NiO_3$ has a pseudocubic (pc) structure with a lattice parameter of $a_{pc}$= 3.814 Å, estimated by averaging the $a_{pc}$ of five parent compounds ($LaNiO_3$, $PrNiO_3$, $NdNiO_3$, $SmNiO_3$, $EuNiO_3$).[18] Given the lattice mismatch between (5A)$NiO_3$ and the substrates, the films experience different strain states: a compressive strain of -0.39 % on LAO and tensile strains of 1.75 % and 2.72 % on LSAT and STO, respectively (**Figure 1a**). XRD θ-2θ scans around the (002) peaks of the (5A)$NiO_3$ films (**Figure 1b**) reveal clear strain-dependent peak shifts, while the full-range XRD patterns are presented in **Figure S1a**. Direct space maps (DSMs) around the substrate (103) Bragg reflection (**Figure S1b**) confirm that all films are coherently strained to the substrates. The progressive shift in the (002) diffraction

peaks (marked by pink arrows) of (5A)$NiO_3$ films toward higher angle from LAO to STO indicates a decrease in the out-of-plane lattice parameter with increasing tensile strain. Distinct Bragg peaks and thickness oscillation fringes confirmed the high crystalline quality of the films. Atomic force microscopy (AFM) measurements reveal smooth surfaces for all as-grown films, with a root mean square roughness of approximately 0.2 nm (**Figure S2**).

**Figures 1c-1e** show cross-sectional high-angle annular dark field scanning transmission electron microscopy (HAADF-STEM) images of approximately 18 nm thick (5A)$NiO_3$ films grown on LAO, LSAT, and STO, respectively, viewed along the [100] zone axis. All films exhibit coherent epitaxy and atomically sharp interfaces with the underlying substrates. Atomic-resolution energy-dispersive X-ray spectroscopy (EDS) composition maps (**Figure 1f**) acquired from the yellow-boxed region in **Figure 1d** show a largely uniform distribution of A-site cations with no apparent atomic-scale segregation or clustering. The film grown on LAO displays a predominantly perovskite structure throughout the imaged region (**Figure 1c**). In contrast, although the majority of the film grown on LSAT remains the perovskite framework, localized regions exhibit diminished Z-contrast between the A-site and B-site columns relative to the surrounding perovskite matrix. Representative perovskite and defective regions are highlighted in **Figures 1g** and **1h**, corresponding to the blue- and red-boxed regions in **Figure 1d**, respectively. The reduced column contrast and disrupted stacking sequence in **Figure 1h** are consistent with the formation of RP-type faults. Similar but more pronounced defective regions are observed in the film grown on STO (**Figure 1e**), whereas such faults are not apparent in the film grown on LAO (**Figure 1c**). These observations indicate that tensile strain preferentially promotes RP fault formation in (5A)$NiO_3$ thin films, consistent with prior reports on $RENiO_3$ films with RE = La or Nd, where tensile strain readily promotes the formation of RP faults.[25, 27]

To probe how strain influences the electronic and transport properties, we performed Ni L-edge XAS, X-ray photoelectron spectroscopy (XPS) valence band (VB), O K-edge XAS, and temperature-dependent resistivity measurements for the set of approximately 13 nm thick films. **Figure 2a** shows Ni $L_{3,2}$ edges XAS for three films alongside $NdNiO_3$ and NiO references representing $Ni^{3+}$ and $Ni^{2+}$ states, respectively.[28] Because the Ni $L_3$ edge partially overlaps with the La $M_4$ edge, the average Ni oxidation state was evaluated primarily from the Ni $L_2$ edge using linear combination analysis with the $Ni^{3+}$ and $Ni^{2+}$ reference spectra (**Figure S3**). As summarized in **Figure 2b**, the average Ni oxidation state decreases systematically with increasing tensile strain. A consistent trend appears at the Ni $L_3$ edge, where the $Ni^{2+}$ feature (black dashed line in **Figure 2a**) becomes more pronounced for the tensile-strained films. The progressive reduction in the Ni oxidation state under tensile strain is consistent with the structural evolution observed by STEM. As shown in **Figure 1**, the film grown on LAO is predominantly perovskite-like, whereas the films grown on LSAT and STO contain RP-type faults, with a higher defect density under stronger tensile strain. Formation of RP faults provides a local stoichiometry-compensation pathway that can accommodate partial reduction of $Ni^{3+}$ toward $Ni^{2+}$. RP-faults can stabilize $Ni^{2+}$-like environments without requiring oxygen vacancies. However, tensile strain may also favor oxygen-vacancy formation by lowering the associated formation energy,[29] potentially contributing to the observed Ni reduction. Because both mechanisms can influence Ni oxidation states, the measured average Ni valence cannot be directly converted into an oxygen-vacancy concentration.

The corresponding in-plane transport behavior further supports this strain-induced structural and valence evolution. **Figure S4** presents the resistivity (ρ) versus temperature (T) relationships for the set of 13 nm thick films. The film grown on LAO remains metallic over the measured temperature range. The film grown on LSAT exhibits a metal-insulator transition around 180 K, while the film grown on STO is insulating throughout the measured temperature range. At room temperature, ρ increases monotonically with tensile strain. The ρ of the film grown on STO is nearly an order of magnitude higher than that of the film grown on LAO. This increase can be attributed to the combined effects of tensile-strain-stabilized RP faults and $V_O^{\cdot\cdot}$,[24, 29-31] both of which reduce the average Ni valence, disrupt Ni–O–Ni connectivity, narrow the Ni 3d bandwidth, and decrease carrier concentration in nickelates.[32]

To gain insight into the strain-stabilized RP-fault formation observed experimentally, we performed first-principles calculations of the thermodynamic stability of stoichiometric perovskite $(5A)NiO_3$ and RP-type $(5A)Ni_{0.5}O_2$ crystal phases under different epitaxial constraints. The corresponding representative structural models are shown in **Figure 2c**. The perovskite model retains corner-sharing $NiO_6$ octahedral connectivity, whereas the RP-type model introduces rock-salt-like A–O layers that interrupt the perovskite framework and locally break the continuity of the Ni–O–Ni network. **Figure 2d** summarizes the calculated relative energies of the perovskite and RP-type phases as a function of in-plane lattice parameter. For the perovskite phase, the energy minimum occurs near the bulk-like lattice parameter (3.814 Å), and the energy increases under both compressive and tensile strain. In contrast, the RP-type phase becomes increasingly favorable as the in-plane lattice parameter expands. This trend indicates that tensile strain lowers the energetic penalty for forming RP-type motifs. The calculated crossover-like behavior is consistent with the experimental observation that the film grown on LAO, i.e., under slight compressive strain, remains predominantly perovskite-like, whereas the films grown on LSAT and STO, i.e., under tensile strain, exhibit RP-type faults; moreover, the density of RP-type faults is increasing upon transition from LSAT to STO substrates. This strain-dependent phase stability also helps rationalize the Ni oxidation state evolution observed by XAS. RP-type faults introduce rock-salt-like A–O layers that interrupt the corner-sharing $NiO_6$ network, locally reduce Ni–O connectivity, and distort the Ni coordination environment. These structural changes favor a more reduced Ni electronic state relative to the fully connected perovskite lattice, leading to a lower average Ni oxidation state in the tensile-strained films. Thus, the systematic decrease in Ni valence from LAO to LSAT and STO is consistent with the increasing density of RP-type faults observed using STEM.

We next examined how these strain-induced structural changes affect the electronic DOS near the $E_F$. **Figure 3a** compares the XPS VB spectra and the pre-edge feature of the O K-edge XAS spectra for the three films. The XPS VB spectra are nearly identical, indicating that the occupied DOS close to $E_F$ is largely unaffected by strain. In contrast, O K pre-edge exhibits a pronounced strain dependence. Relative to the film grown on LAO, the tensile-strained films grown on LSAT and STO show a shift of the pre-edge feature toward lower photon energy, together with enhanced intensity. The full O K-edge spectra are shown in **Figure S5**, confirming that the most pronounced strain-dependent changes occur in the pre-edge region associated with unoccupied Ni 3d–O 2p hybridized states. This behavior indicates that the unoccupied DOS near $E_F$ is strongly modified by tensile strain.

At first glance, this trend appears counterintuitive. A lower average Ni oxidation state would normally be expected to reduce the density of ligand-hole states and shift the O K pre-edge to higher energy.[28, 33, 34] Instead, we observe a lower-energy and more intense pre-edge under tensile strain. This suggests that the O K pre-edge evolution cannot be explained solely by changes in Ni oxidation state and instead points to a strain-induced redistribution of low-energy unoccupied states. The enhanced pre-edge intensity implies that an additional strain-activated contribution to the low-energy unoccupied DOS is involved.

To quantify this effect, we calculated the projected DOS for both perovskite and RP-type $(5A)NiO_3$-derived structures as a function of in-plane lattice parameters. **Figures 3b,c** compare the momentum-resolved DOS strain-dependent evolution for the perovskite and RP-type structures. A complementary DOS representation, shown in **Figure S6**, provides the same comparison with the total DOS overlaid with p-, d-, and f-orbital contributions, highlighting the strain-induced shift of low-energy unoccupied states. For the stoichiometric perovskite model (**Figure 3b**), tensile strain shifts the unoccupied states toward lower energy. The projected DOS shows that this shift is dominated by RE 4f-derived states, which move closer to $E_F$ under tensile strain and provide an additional reservoir of low-energy unoccupied states. Element-resolved 4f-projected DOS further confirms that tensile strain shifts the unoccupied 4f states of the rare-earth cations toward lower energy in the perovskite phase, with the strongest shift observed for Pr (**Figure S7**). A similar strain-dependent shift is observed in the RP-type model (**Figure 3c**). Rare-earth-resolved analysis for the RP-type structure also shows tensile-strain-induced lowering of unoccupied 4f states, with the strongest shifts observed for Eu and Sm (**Figure S8**). In addition, we performed additional calculations to examine the effect of U values applied separately for the 4f states of lanthanides and for 3d states of Ni, defined as U(RE) and U(Ni), respectively. The resulting 4f-projected DOS for U(RE) and U(Ni) values in the 2–4 eV range are shown in **Figure S9**. Representative three sets of U values were selected to illustrate the case U(RE)=U(Ni) (gray), U(RE)>U(Ni) (pink), and U(RE)<U(Ni) (blue). For the U(RE)=U(Ni) and U(RE)>U(Ni) conditions, increasing U(RE) from 2 eV to 3 eV and then 4 eV, the shift of the lowest energy peak (indicated with a dotted line) remains approximately the same. However, the magnitude of this peak changes significantly, resulting in DOS depletion under compressive epitaxial strain. In other words, tensile strain shifts the RE 4f states toward the Fermi level, in agreement with our experimental results.

These calculations support the schematic picture shown in **Figure 3d**. Under compressive strain, the low-energy unoccupied DOS near $E_F$ is dominated primarily by Ni 3d–O 2p hybridized states, with RE 4f levels located farther from $E_F$. Under tensile strain, lattice expansion narrows the Ni 3d bandwidth while drawing RE 4f states closer to $E_F$. The resulting RE 4f/O 2p/Ni 3d hybridization introduces additional low-energy unoccupied states, consistent with the intensified O K pre-edge observed for the tensile-strained films. Although the rare-earth 4f-derived states remain above the Fermi level, their strain-dependent energetic alignment may modify Ni–O electronic interactions through coupling with O 2p and Ni 3d states.[35] Such changes could influence orbital interactions and charge redistribution between surface Ni–O sites and adsorbed OER intermediates (OH*, O*, and OOH*), potentially affecting their bonding strengths and adsorption energetics. However, establishing a quantitative relationship would require explicit surface adsorption calculations, which are beyond the scope of the present study.

For the film on STO, stronger tensile strain further promotes Ni reduction and RP-fault formation, which can narrow the Ni 3d bandwidth and partially offset the beneficial increase in low-energy unoccupied DOS. Thus, the film on LSAT represents an intermediate tensile-strain regime where the unoccupied DOS is enhanced without excessive disruption of the Ni–O network. Taken together, the combined experimental and theoretical results reveal that epitaxial strain plays two coupled roles in A-site high-entropy nickelate films. First, tensile strain stabilizes RP-type faults that interrupt the perovskite framework, leading to partial Ni reduction and increased defect density. Second, tensile strain modifies the unoccupied electronic structure by shifting RE 4f-derived states toward $E_F$, thereby enhancing RE 4f/O 2p/Ni 3d hybridization.

We then investigated how epitaxial strain influences the OER activity of $(5A)NiO_3$ thin films. The 13 nm thick films were electrically contacted on the front and tested in $O_2$-saturated 0.1 mol/L KOH using a conventional three-electrode cell.[7, 12, 13] The first-cycle iR-corrected cyclic voltammetry (CV) curves (**Figure 4a**) reveal a clear strain dependence. The compressively strained film on LAO shows the latest OER onset and the lowest current density, whereas the tensile-strained films on LSAT and STO exhibit earlier catalytic onset and substantially higher current densities. The inset of **Figure 4a** highlights the precatalytic Ni redox region. The films on LSAT and STO show pronounced Ni redox features, while the film on LAO exhibits a much weaker response. These redox peaks are commonly associated with the oxidation of $Ni^{2+}$ to higher-valent $Ni^{3+}/Ni^{4+}$ species during electrochemical polarization.[7, 36]

The strain-dependent activity is summarized in **Figure 4b** by comparing the current density at 1.63 V vs RHE. In the first cycle, the film on LSAT exhibits the highest activity, followed by the film on STO, and then the film on LAO. This trend persists after five cycles, although the film on LSAT exhibits a moderate decrease in current density (**Figure S10**). In contrast, the film on STO shows relatively stable activity over the first five cycles, whereas the film on LAO undergoes gradual activation. The nonmonotonic activity trend suggests that moderate tensile strain provides the most favorable balance between competing structural and electronic effects. Tensile strain promotes RP-type fault formation and partial Ni reduction, increasing the population of $Ni^{2+}$-like sites that participate in pre-catalytic redox processes, consistent with the stronger Ni redox features observed for the films on LSAT and STO (**Figure 4a**). The formation of oxygen vacancies is another means to accommodate the lower Ni valence, affecting redox activity and electronic transport, although their independent contributions cannot be determined from the present measurements. However, excessive RP-faults and/or oxygen vacancies can disrupt the corner-sharing $NiO_6$ network, reduce Ni–O–Ni connectivity, and hinder electronic transport. This competing effect is consistent with the lower activity of the film on STO relative to the film on LSAT, despite its greater RP-fault density and stronger Ni reduction. Meanwhile, tensile strain shifts rare-earth 4f-derived states toward the Fermi level, modifying the low-energy unoccupied DOS and RE 4f/O 2p/Ni 3d hybridization (**Figure 3**). These electronic changes alter the bonding environment of Ni–O sites and are believed to favor the OER.

Unlike B-site high-entropy perovskites, where multiple transition-metal cations can directly participate in catalytic redox processes,[37] A-site compositional complexity in (5A)$NiO_3$ offers a distinct approach to tuning the Ni-based catalytic environment. Incorporating five rare-earth elements introduces diverse local chemical environments that influence lattice distortion and defect formation. Beyond changes in the initial Ni oxidation state, our O K-edge XAS measurements and first-principles calculations (**Figure 3**) reveal that strain-dependent changes in unoccupied electronic states, including contributions from rare-earth 4f-derived states that may influence Ni–O electronic interactions. Thus, A-site compositional complexity provides additional structural and electronic degrees of freedom for tuning Ni-based OER catalysts, although their individual contributions to catalytic kinetics remain to be established.

To benchmark the catalytic performance, we compare our (5A)$NiO_3$ films with representative $LaNiO_3$ and $NdNiO_3$ thin film electrocatalysts reported in the literature (**Figure 4b**).[8, 23-25, 38] Both the first- and fifth-cycle activities of tensile-strained (5A)$NiO_3$ films exceed most reported single-A-site nickelate films at comparable strain states. This comparison highlights the distinct role of A-site high-entropy disorder: while the active B-site remains Ni, the mixed rare-earth A-site sublattice introduces additional structural and electronic degrees of freedom, including strain-tunable RP-fault formation and RE 4f-derived unoccupied states. These effects contribute to high OER activity of tensile-strained (5A)$NiO_3$ films relative to conventional single-A-site nickelate thin films.

We also compared the OER performance of our approximately 13 nm thick (5A)$NiO_3$/LSAT film with the 30 and 45 unit-cell-thick (5A)$NiO_3$/$NdGaO_3$ films reported by Patel et al.,[39] which have comparable strain states and thicknesses. At the selected potential, the OER current densities reported by Patel et al. are approximately 100 μA $cm^{-2}$,[39] lower than the value (>400 μA $cm^{-2}$) observed for our (5A)$NiO_3$/LSAT film. One possible contributing factor is oxygen nonstoichiometry. Our films were post-annealed under 75 Torr $O_2$, compared with 500 Torr $O_2$ in the study by Patel et al.[39] The lower annealing oxygen pressure, together with tensile strain, may favor oxygen-vacancy formation and influence Ni oxidation states and electrochemical behavior.[38]

In addition to activity, structural stability under OER conditions is a key requirement for oxide electrocatalysts.[3, 40, 41] We therefore compared the long-term electrochemical response and post-OER structural evolution of the most active high-entropy nickelate film, (5A)$NiO_3$/LSAT, with a representative $LaNiO_3$/LSAT film. **Figure 5a** shows chronoamperometry (CA) curves collected at 1.64 V vs RHE in $O_2$-saturated 0.1 mol/L KOH for 15 h. The $LaNiO_3$ film exhibits a pronounced initial decrease in current density during the first several hours, followed by a gradual recovery to a value slightly above its initial current. In contrast, the (5A)$NiO_3$ film shows a more gradual current-density evolution, reaching a modest increase after 15 h. These distinct responses suggest different degrees or pathways of electrochemical surface activation and surface reconstruction, with the high-entropy film undergoing a less pronounced change in catalytic current.

Ex situ XRD further confirms that both films undergo OER-induced structural modifications, but to different extents (**Figure 5b**). For (5A)$NiO_3$/LSAT, the primary perovskite diffraction peak is preserved after OER, and fitting of the Laue fringes indicates that the coherently diffracting crystalline thickness decreases from approximately 13 nm to 11 nm. This reduction is consistent with the formation of a thin, amorphous and/or structurally reconstructed surface layer.[42, 43] In contrast, the coherently diffracting thickness of $LaNiO_3$/LSAT decreases from approximately 13 nm to 8.5 nm, indicating a thicker amorphous or structurally reconstructed region and/or greater material loss. Thus, although both films undergo surface reconstruction during OER, (5A)$NiO_3$ retains a larger fraction of its crystalline lattice after prolonged operation.

To further investigate the structural and chemical evolution of (5A)$NiO_3$ as a result of OER operation, we performed post-OER AFM and cross-sectional STEM-EDS characterization (**Figure S11**). AFM reveals changes in surface morphology after OER, while STEM-EDS elemental mapping and line-profile analysis show relatively uniform distributions of the five rare-earth cations throughout the film interior, with approximately equimolar A-site proportions and no pronounced element-specific segregation. However, within the top ~2.5 nm, the measured Ni atomic fraction decreases noticeably, accompanied by a smaller reduction in the rare-earth cation fractions, indicating changes in the near-surface composition. Such changes may arise from OER-induced surface reconstruction,[43, 44] during which the underlying perovskite framework can serve as a support for transition-metal (oxy)hydroxide active species, potentially accompanied by surface passivation or amorphization.[45] Together with the post-OER XRD results, these observations indicate that (5A)$NiO_3$ largely preserves its underlying crystalline framework and multicomponent A-site composition despite near-surface modifications. Further investigation is needed to distinguish the specific contributions of surface reconstruction, cation dissolution, and amorphization.

## 3. Conclusion

In this study, we investigated how epitaxial strain influences defect formation, Ni oxidation state, electronic structure, and OER performance in A-site high-entropy (5A)$NiO_3$ thin films. Atomic-resolution STEM reveals that tensile strain promotes the formation of RP-type faults, which introduce rock-salt-like A–O layers into the perovskite framework, interrupt the corner-sharing $NiO_6$ network, and favor more $Ni^{2+}$-like character. Consistently, Ni L-edge XAS reveals a systematic decrease in the average Ni oxidation state with increasing tensile strain. While the occupied density of states near the $E_F$ remain largely unchanged, the O K-edge pre-edge shifts to lower photon energy and increases in intensity under tensile strain, indicating a redistribution of low-energy unoccupied states. First-principles calculations suggest that tensile strain shifts rare-earth 4f-derived states toward $E_F$, modifying the electronic interactions involving RE 4f, O 2p, and Ni 3d orbitals. Although these calculated shifts are relatively modest, they reveal an additional strain-sensitive contribution to the unoccupied electronic structure associated with the chemically diverse A-site sublattice, which may influence Ni–O electronic interactions and the bonding environment of catalytic sites. Thus, beyond the previously established roles of strain-induced RP-fault formation and Ni oxidation-state changes in conventional nickelates,[24, 25, 27] our findings highlight the importance of considering A-site-derived electronic states when interpreting the properties of compositionally complex nickelates.

The tensile-strained films on LSAT and STO exhibit earlier OER onset, stronger pre-catalytic $Ni^{2+} \rightarrow Ni^{3+}/Ni^{4+}$ redox features, and substantially higher current densities than the compressively strained film on LAO. Among them, the film under moderate tensile strain on LSAT shows the highest activity, highlighting the competing effects of strain-induced structural and electronic changes. While RP-fault formation and partial Ni reduction may introduce additional redox-active environments, excessive disruption of the Ni–O network can hinder electronic transport and offset these benefits. Importantly, the calculated DOS does not directly quantify OER kinetics, and the substantial activity enhancement cannot be attributed to the RE 4f-state shifts alone. Instead, their potential contribution must be considered alongside strain-induced RP-fault formation and oxygen vacancy formation. Beyond catalytic activity, $(5A)NiO_3$ exhibits a more stable electrochemical response and a smaller reduction in coherently diffracting crystalline thickness than $LaNiO_3$ after 15 h of OER operation, indicating improved resistance to OER-induced structural degradation.

Overall, by revealing the strain-dependent evolution of rare-earth 4f-derived states alongside structural defect formation and their potential contributions to OER behavior, our findings provide new insights into the interplay among A-site chemistry, defect structures, and electronic interactions in high-entropy oxide electrocatalysts. This work highlights A-site compositional complexity as an additional structural and electronic design parameter for Ni-based oxide electrocatalysts, extending beyond conventional approaches focused on epitaxial strain and transition-metal oxidation states.

## 4. Experimental Section/Methods

*Film growth*: Epitaxial (5A)$NiO_3$ thin films were deposited on (001)-oriented LAO, LSAT, and STO single-crystal substrates using PLD (TSST). STO substrates were etched for 30 s in buffered HF and annealed in air at 1000 °C for 1 h to achieve the $TiO_2$-terminated surface. Prior to deposition, the substrates were sequentially ultrasonicated in acetone and isopropanol (5 min each) and dried with nitrogen gas. During the deposition, a KrF excimer laser ($\lambda$ = 248 nm) was operated at a laser fluence of 1.5 J cm$^{-2}$ and a repetition rate of 3 Hz. The substrate temperature was maintained at 700 °C, and the oxygen partial pressure was set at 13 Pa (100 mTorr). After deposition, the films were cooled to 500 °C and annealed for 30 min under the oxygen pressure of 1 kPa (75 Torr), then cooled down to room temperature in 1 kPa (75 Torr) $O_2$.

*Characterization*: Crystal structure was characterized using high-resolution X-ray diffraction (Rigaku SmartLab) with Cu Kα radiation (l = 1.5406 Å). r-T curves of the samples were investigated using a Hall measurement system (Ecopia) in the temperature range 90 – 310 K. Measurements were done in the van der Pauw geometry with square samples (approximately 5 mm × 5 mm) and silver paste in the corners. Surface morphology was examined by AFM (Bruker MultiMode 8) in tapping mode using a ATEC-FM (approximately 2 N/m) tip. Cross-sectional lamellae for TEM were prepared using a Thermo Fisher Hydra plasma focused ion beam scanning electron microscope equipped with an Xe ion source. HAADF STEM images were acquired on an aberration-corrected Thermo Fisher Spectra Ultra S/TEM operated at 300 kV with a 42 pA probe current. The convergence angle was set to 30 mrad and the collection angle ranged from 68 to 180 mrad. XPS VB measurements were performed at the National Institute of Standards and Technology operated beamline 7-ID-2 of the National Synchrotron Light Source II (NSLS-II) using 2 keV X-rays. Ni L-edge and O K-edge soft XAS measurements were conducted at NSLS-II beamline 23-ID-2 (IOS), both in total electron yield mode through drain current measurements and partial fluorescence yield mode using a Vortex EM silicon drift detector.

*Ab initio simulations:* Strained epitaxial films of (5A)$NiO_3$ perovskite and (5A)$Ni_{0.5}O_2$ Ruddlesden-Popper (RP) phases were represented using the periodic model approach and supercells formed by the 2×2×4 extensions of the corresponding crystallographic cells $A_{16}Ni_{16}O_{48}$ and $A_{16}Ni_8O_{32}$, respectively. The 5A composition of the A-sites was mimicked by randomly distributed La, Pr, Nd, Sm, and Eu atoms at the ratio of 3:3:3:3:4. For comparison, we also considered pure perovskite and RP compounds as discussed in the text.

The calculations were carried out within the density functional theory (DFT) formalism implemented in the Vienna ab initio Simulation Package (VASP).[46,47] The projector-augmented wave method was used to represent core electrons.[48] Exchange–correlation interactions were described using the functional by Perdew-Burke-Ernzerhof (PBE).[49] Kohn–Sham states were expanded in a plane-wave basis with a 500 eV cutoff; the calculations were performed using the 2×2×1 k-mesh. The rotationally invariant +U correction[50] was applied to the 4f states of all lanthanides and the 3d states of Ni. For simplicity, we used the smallest value (U = 2 eV) that ensured convergence of the spin-polarized solution throughout. To further examine the effect of U values applied separately for the 4f states of lanthanides and for 3d states of Ni, we performed additional calculations in which U was varied independently from 2 eV to 4 eV for the lanthanide 4f states and the Ni 3d states. The $a^{+}b^{-}c^{-}$ Glazner tilt system was used in all cases unless otherwise noted. All calculations were performed in the spin-polarized mode in the ferromagnetic configuration of the A-site and Ni spins. The energies of all structures were optimized until the total energy change was below $10^{-5}$ eV. The population analysis was performed using the Bader method.[51]

The in-plane lattice parameters were fixed at the corresponding experimental values for SLAO (3.756 Å), LAO (3.794 Å), LSAT (3.868 Å), and STO (3.905 Å). Since PBE functional tends to overestimate lattice parameters by approximately 1%, we simulated the experimental tensile conditions using the lattice parameters 1% larger than those of STO (3.944 Å), denoted as "STO+1%". The out-of-plane parameter was optimized for every stoichiometric and RP phase configuration.

*Electrochemical Experiments*: Electrical contact to the films was made by applying Ga-In eutectic (Sigma-Aldrich, >99.99%) to a scratched corner of each film, attaching a Cu wire with Ag paste (Liebsilber, Ted Pella), and insulating the back and sides—including the wire contact—with chemically inert epoxy (Loctite 9460), leaving only the film surface exposed to the electrolyte. 0.1 mol/L KOH electrolyte (pH 13) was prepared by dissolving high-purity KOH pellets (Sigma-Aldrich, 99.99%) in Nanopure water (18.2 MΩ cm). Approximately 15 mL of electrolyte was stirred and bubbled with ultra-high-purity oxygen gas before and during the measurements. A Pt wire and a saturated calomel electrode (RE-2BP, ALS Co.) were used as the counter and reference electrodes, respectively. All electrochemical experiments were conducted using a Biologic SP-200 potentiostat.

**Supporting Information**

Supporting Information is available from the Wiley Online Library or from the author.

**Acknowledgements**

This work was mainly supported by the U.S. Department of Energy (DOE), Office of Science (SC), Office of Basic Energy Sciences, Division of Materials Sciences and Engineering, Synthesis and Processing Science Program, under award No. 10122. Partial data analysis (including STEM images and spectroscopy) and manuscript preparation were supported by the U.S. DOE, SC, Early Career Research Program under Award No. 85673. This research used resources of the 7-ID-2 (SST-2) and 23-ID-2 (IOS) beamlines of the National Synchrotron Light Source II, a U.S. Department of Energy (DOE) Office of Science User Facility operated for the DOE Office of Science by Brookhaven National Laboratory under Contract No. DE-SC0012704. This research used resources of the National Energy Research Scientific Computing Center, a DOE SC User Facility supported by the Office of Science of the U.S. Department of Energy under Contract No. DE-AC02-05CH11231 using NERSC award BES-ERCAP0037250. NIST work was funded solely by the United States Government. The work at Los Alamos National Laboratory was supported by the NNSA's Laboratory Directed Research and Development Program, and was performed, in part, at the CINT, an Office of Science User Facility operated for the U.S. Department of Energy, Office of Science. Los Alamos National Laboratory, an affirmative action equal opportunity employer, is managed by Triad National Security, LLC for the U.S. Department of Energy's NNSA, under contract 89233218CNA000001.

**Conflict of Interest**

The authors declare no conflict of interest.

**Data Availability Statement**

The data that support the findings of this study are available from the corresponding author upon reasonable request.

**Disclaimer**

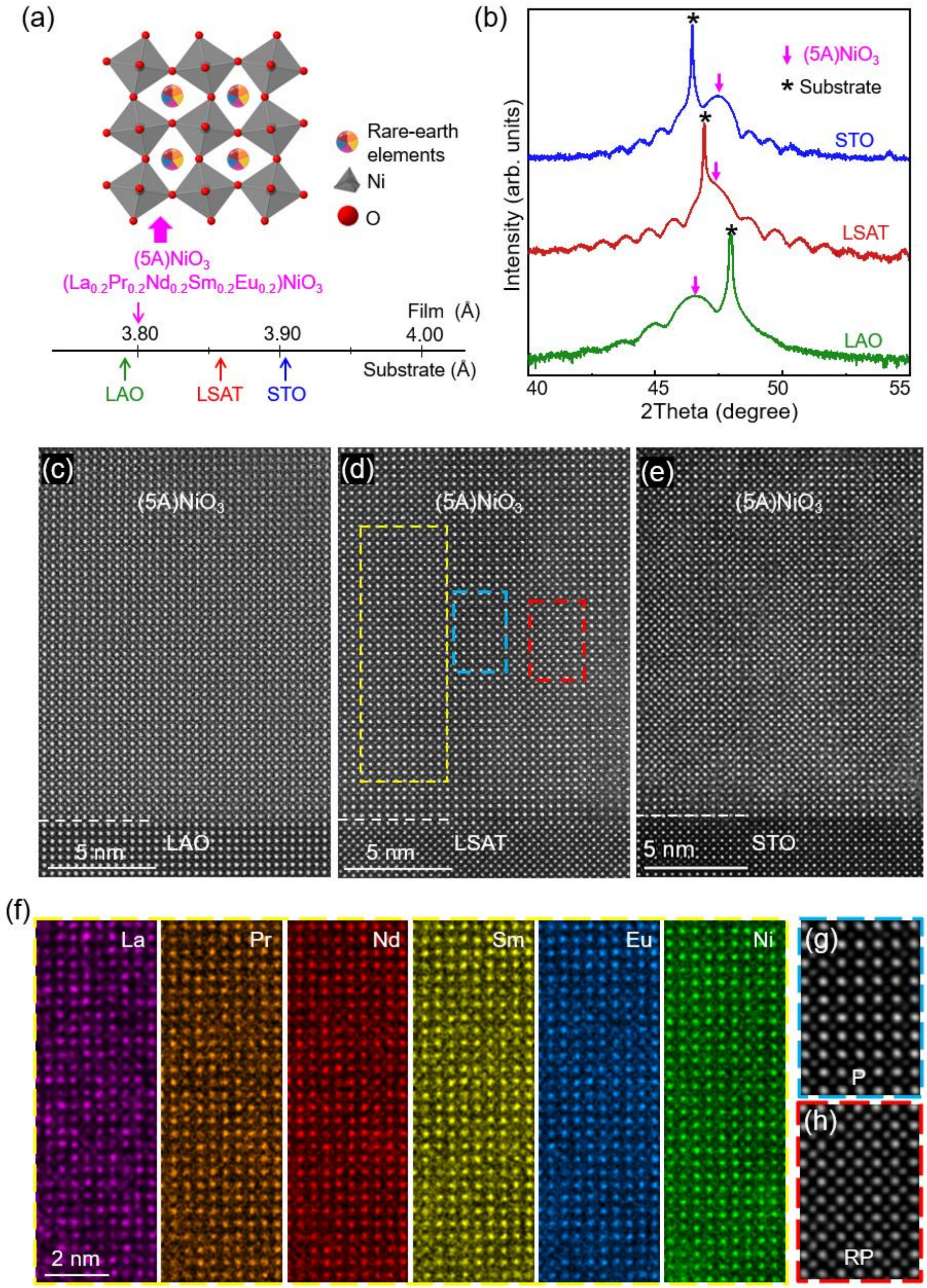

**Figure 1. Strain-dependent structural evolution and RP-fault formation in epitaxial $(5A)NiO_3$ thin films.** (a) Schematic crystal structure of A-site high-entropy $(5A)NiO_3$, where 5A = $(La_{0.2}Pr_{0.2}Nd_{0.2}Sm_{0.2}Eu_{0.2})$ together with the lattice parameters of bulk $(5A)NiO_3$ and the LAO, LSAT, and STO substrates. (b) XRD θ-2θ scans around the (002) diffraction peak. Pink arrows mark the $(5A)NiO_3$ film peaks, and asterisks denote substrate peaks. (c,d) Cross-sectional HAADF-STEM images of $(5A)NiO_3$ films grown on (c) LAO, (d) LSAT, and (e) STO. (f) Atomic-resolution STEM-EDS elemental maps acquired from the yellow-boxed region in (d), showing a largely uniform distribution of La, Pr, Nd, Sm, Eu, and Ni. (g,h) Enlarged HAADF-STEM images from the blue- and red-boxed regions in (d), respectively, highlighting a defect-free perovskite region (P) and an RP-fault-containing region (RP).

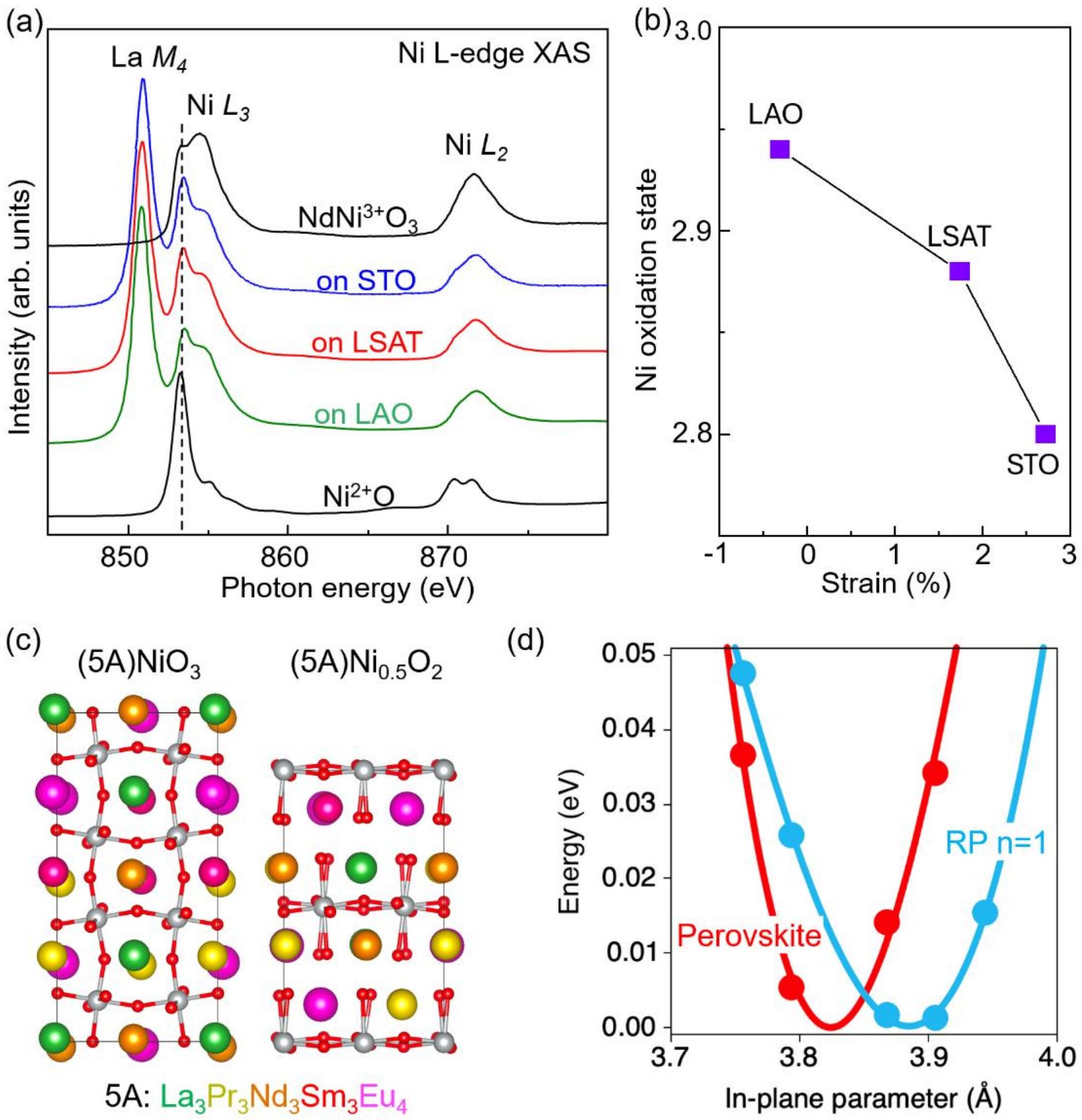

**Figure 2. Strain-dependent Ni oxidation state and calculated stability of perovskite and RP-type motifs in (5A)$NiO_3$ thin films.** (a) Ni $L_{3,2}$-edge XAS spectra of (5A)$NiO_3$ films grown on LAO, LSAT, and STO, together with $NdNiO_3$ and NiO references representing $Ni^{3+}$ and $Ni^{2+}$ states, respectively. The dashed line marks the $Ni^{2+}$-related feature near the Ni $L_3$ edge. (b) Average Ni oxidation state extracted from Ni $L_2$-edge linear-combination analysis as a function of epitaxial strain. (c) Structural models used in first-principles calculations for stoichiometric perovskite (5A)$NiO_3$ and RP-type (5A)$Ni_{0.5}O_2$ motifs. (d) Calculated relative energies of the perovskite and RP-type structures as a function of in-plane lattice parameter, showing that tensile strain lowers the energetic penalty for forming RP-type motifs.

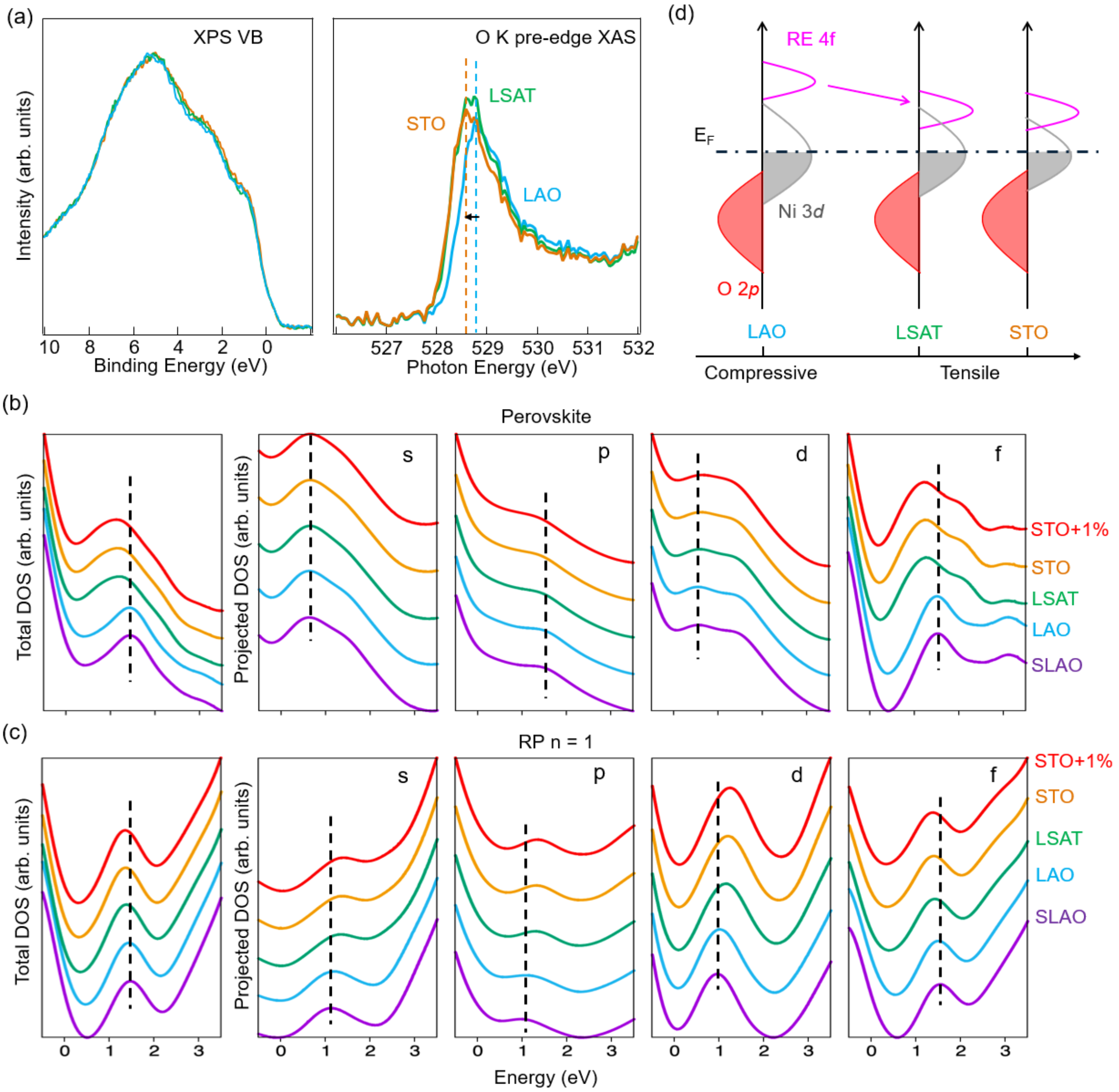

**Figure 3. Strain-modulated occupied and unoccupied electronic structures of (5A)$NiO_3$ thin films.** (a) XPS valence-band spectra (left) and O K-edge pre-edge XAS spectra (right) of (5A)$NiO_3$ films grown on LAO, LSAT, and STO. (b,c) Calculated total and projected DOS for (b) stoichiometric perovskite and (c) RP-type structures under different in-plane lattice parameters corresponding to compressive and tensile strain states. Orbital-resolved contributions from total, s, p, d, and f states are shown. Dashed lines indicate the strain-dependent shift of representative unoccupied states toward lower energy; the Fermi energy ($E_F$) is located at 0 eV. (d) Schematic electronic-structure evolution from compressive to tensile strain. Tensile strain draws RE 4f-derived states closer to $E_F$.

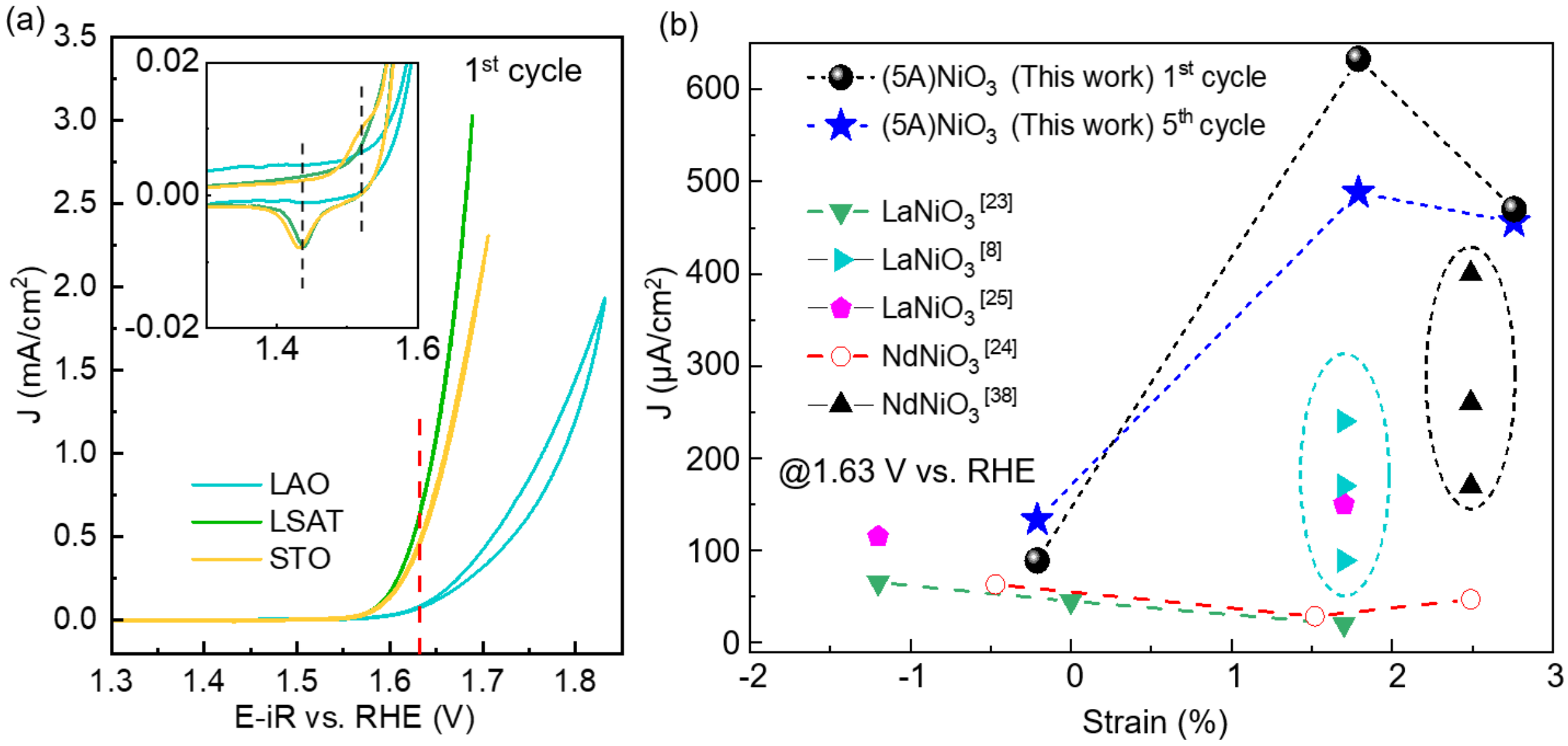

**Figure 4. Strain-dependent OER activity of (5A)$NiO_3$ thin films.** (a) First-cycle iR-corrected CV curves of (5A)$NiO_3$ films grown on LAO, LSAT, and STO, measured in $O_2$-saturated 0.1 mol/L KOH. The inset highlights the pre-catalytic Ni redox region, where tensile-strained films show more pronounced Ni redox features than the compressively strained LAO film. (b) Comparison of OER activity, expressed as current density at 1.63 V vs RHE, as a function of epitaxial strain. The first- and fifth-cycle activities of (5A)$NiO_3$ are compared with representative $LaNiO_3$ and $NdNiO_3$ thin-film electrocatalysts reported in the literature.[8, 23-25, 38] Tensile-strained (5A)$NiO_3$ films exhibit substantially enhanced activity, with the LSAT film showing the highest current density.

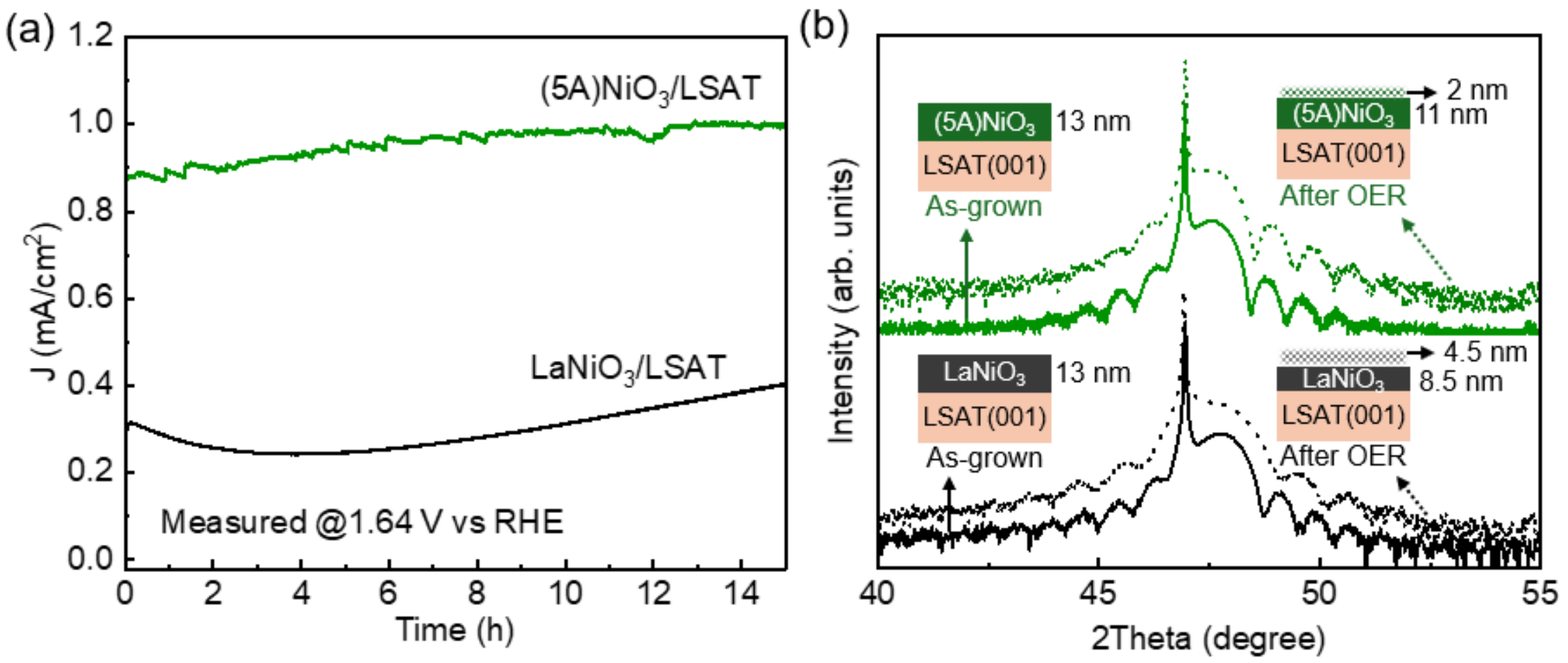

**Figure 5. Long-term stability and surface evolution of (5A)$NiO_3$ and $LaNiO_3$ thin films.** (a) Chronoamperometry curve of the (5A)$NiO_3$/LSAT and $LaNiO_3$/LSAT films measured at a constant overpotential of 410 mV (1.64 V vs RHE) for 15 h in $O_2$-saturated 0.1 mol/L KOH. (b) Out-of-plane XRD θ-2θ scans of (5A)$NiO_3$ and $LaNiO_3$ films on LSAT before and after OER stability testing. The primary perovskite diffraction peak is preserved after OER, while changes in Laue-fringe spacing indicate a reduction in coherently diffracting crystalline thickness, consistent with the formation of a reconstructed surface region.

**High-Entropy A-Site Disorder Enhances Oxygen Evolution Reaction Activity and Stability in Nickelate Thin Films**

*In high-entropy ($La_{0.2}Pr_{0.2}Nd_{0.2}Sm_{0.2}Eu_{0.2}$)$NiO_3$ thin films, tensile strain promotes Ruddlesden–Popper fault formation, lowers Ni oxidation state, and shifts rare-earth 4f-derived states toward the Fermi level, strengthening rare-earth 4f/O 2p/Ni 3d hybridization. These effects enhance OER activity beyond representative single-A-site nickelate films while improving structural stability, demonstrating a design strategy coupling strain and A-site disorder for robust oxide electrocatalysts.*

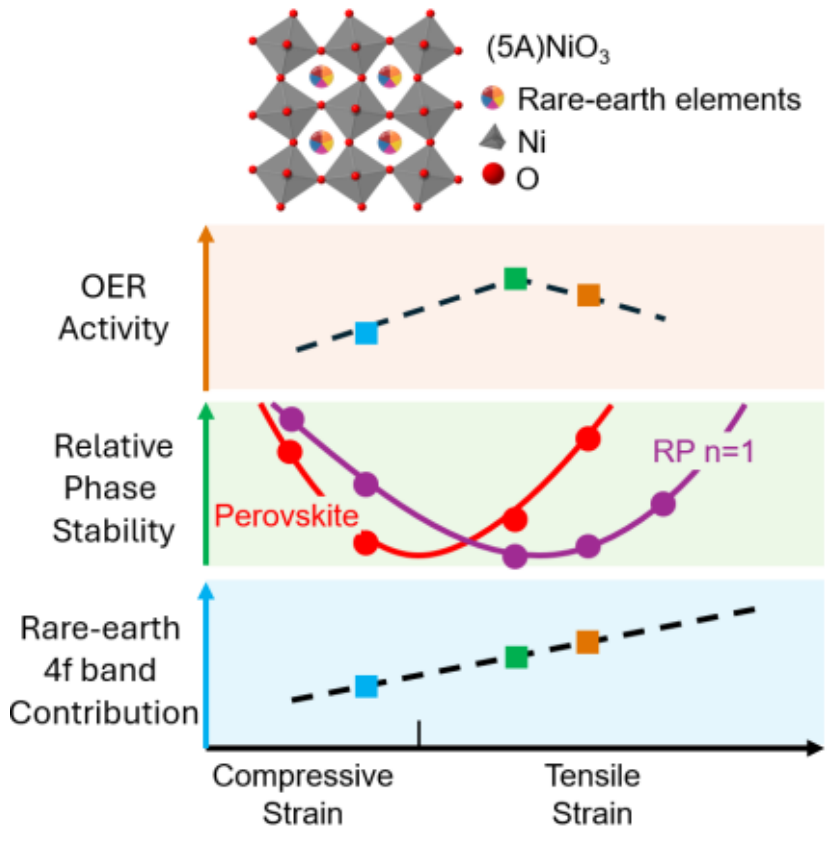

# Supporting Information

**High-Entropy A-Site Disorder Enhances Oxygen Evolution Reaction Activity and Stability in Nickelate Thin Films**

*Min-Ju Choi, Le Wang*, Krishna Prasad Koirala, Hyoju Park, Mark E. Bowden, Elizabeth Helen Krenkel, Aiping Chen, Jueli Shi, Conan Weiland, Iradwikanari Waluyo, Adrian Hunt, Peter V. Sushko, and Yingge Du**

Dr. M. Choi, Dr. L. Wang, Dr. K. P. Koirala, Dr. H. Park, Dr. M. E. Bowden, Dr. J. Shi, Dr. P. V. Sushko, Dr. Y. Du
Integrated Discovery Sciences Directorate, Pacific Northwest National Laboratory, Richland, Washington 99354, United States
E-mail: le.wang@pnnl.gov, yingge.du@pnnl.gov
Dr. K.P. Koirala
Materials Department, University of California Santa Barbara, Santa Barbara, California 93106, United States
Dr. E. H. Krenkel, Dr. A. Chen
Center for Integrated Nanotechnologies, Los Alamos National Laboratory, Los Alamos, New Mexico 87545, United States
Dr. C. Weiland
Material Measurement Laboratory, National Institute of Standards and Technology, Gaitherburg, Maryland 20899, United States
Dr. I. Waluyo, Dr. A. Hunt
National Synchrotron Light Source II, Brookhaven National Laboratory, Upton, New York 11973, United States

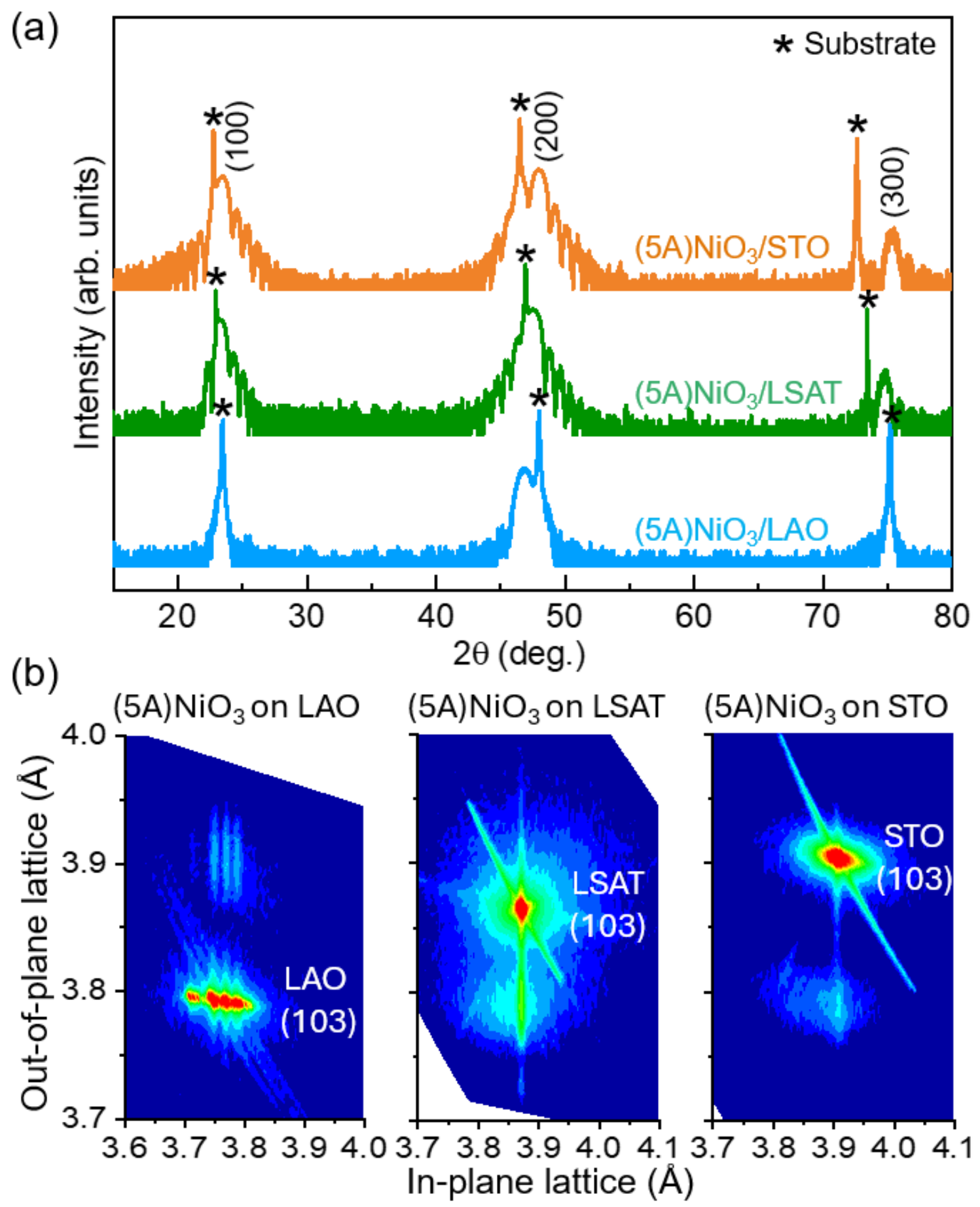

**Figure S1.** (a) Full range XRD $\theta$-$2\theta$ scans for the $(5A)NiO_3$ thin films epitaxially grown on (001)-oriented LAO, LSAT, and STO single-crystalline substrates. (b) Direct space mapping (DSMs) for $(5A)NiO_3$ films on LAO, LSAT, and STO substrates.

The relatively weak (001) and (003) diffraction intensities of $(5A)NiO_3$ grown on LAO can be attributed to a combination of reduced film thickness and substrate-related diffraction effects. The film on LAO is approximately 11 nm thick, compared with approximately 13 nm for the films on LSAT and STO, resulting in lower overall diffraction intensity. In addition, the film (001) reflection lies close to the LAO substrate peak, making the relatively weak film peak difficult to resolve, but it can be seen as a shoulder in **Figure S1a**. Furthermore, LAO has a rhombohedral crystal structure at room temperature and exhibits ferroelastic twin domains, as evidence in the direct space map of this sample (**Figure S1b**). At small diffraction angles, the peaks from different twin domains are close enough to be combined in our θ–2θ scan. However, at the large angle of the (003) peak, the angular separation between twin-domain reflections becomes larger, such that only a portion of the diffracted intensity may fall within the measurement's angular acceptance. Combined with the relatively smaller film thickness, this effect likely accounts for the suppression of the (003) reflection to near-background levels.

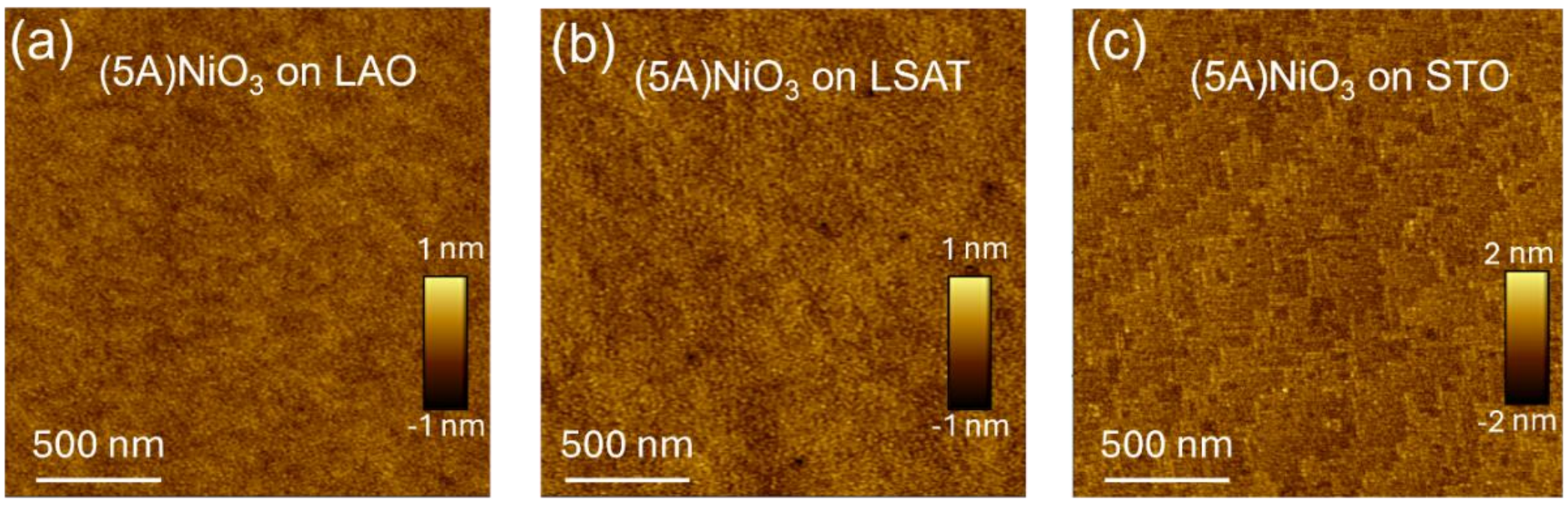

**Figure S2.** AFM images of (5A)$NiO_3$ films grown on LAO (a), LSAT (b), and STO (c).

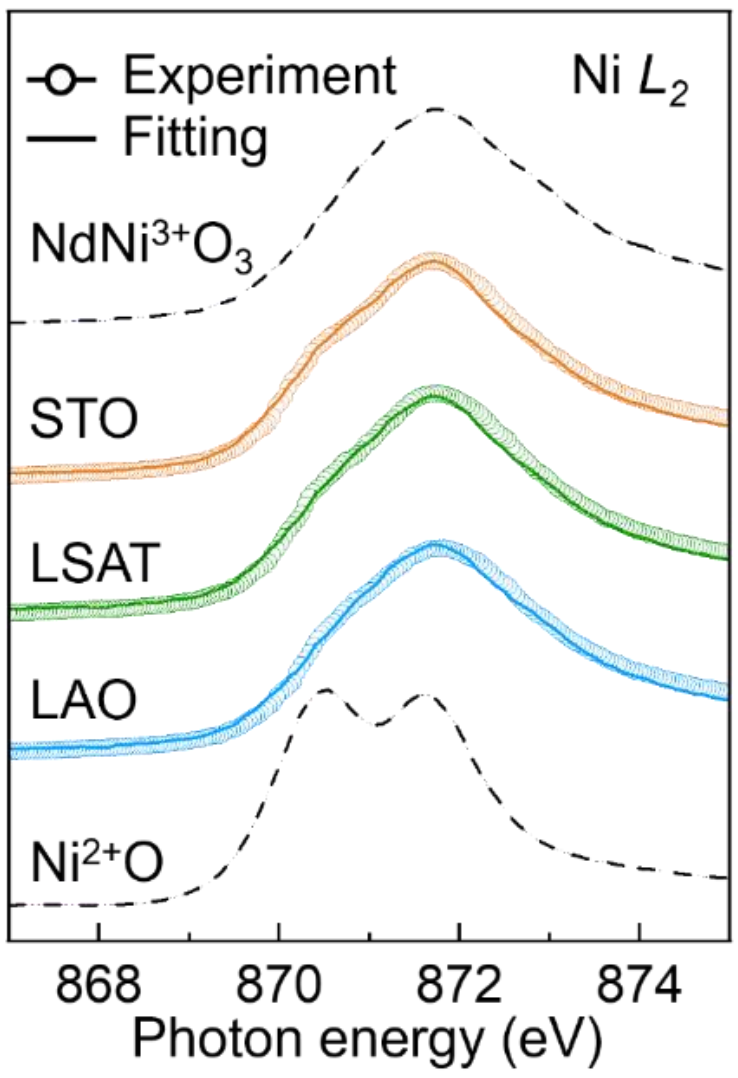

**Figure S3.** Ni $L_2$-edge XAS spectra of $(5A)NiO_3$ films grown on LAO, LSAT, and STO, shown together with $NdNiO_3$ and NiO reference spectra representing $Ni^{3+}$ and $Ni^{2+}$ states, respectively. Open circles denote the experimental spectra, and solid lines represent linear-combination fits used to estimate the average Ni oxidation state.

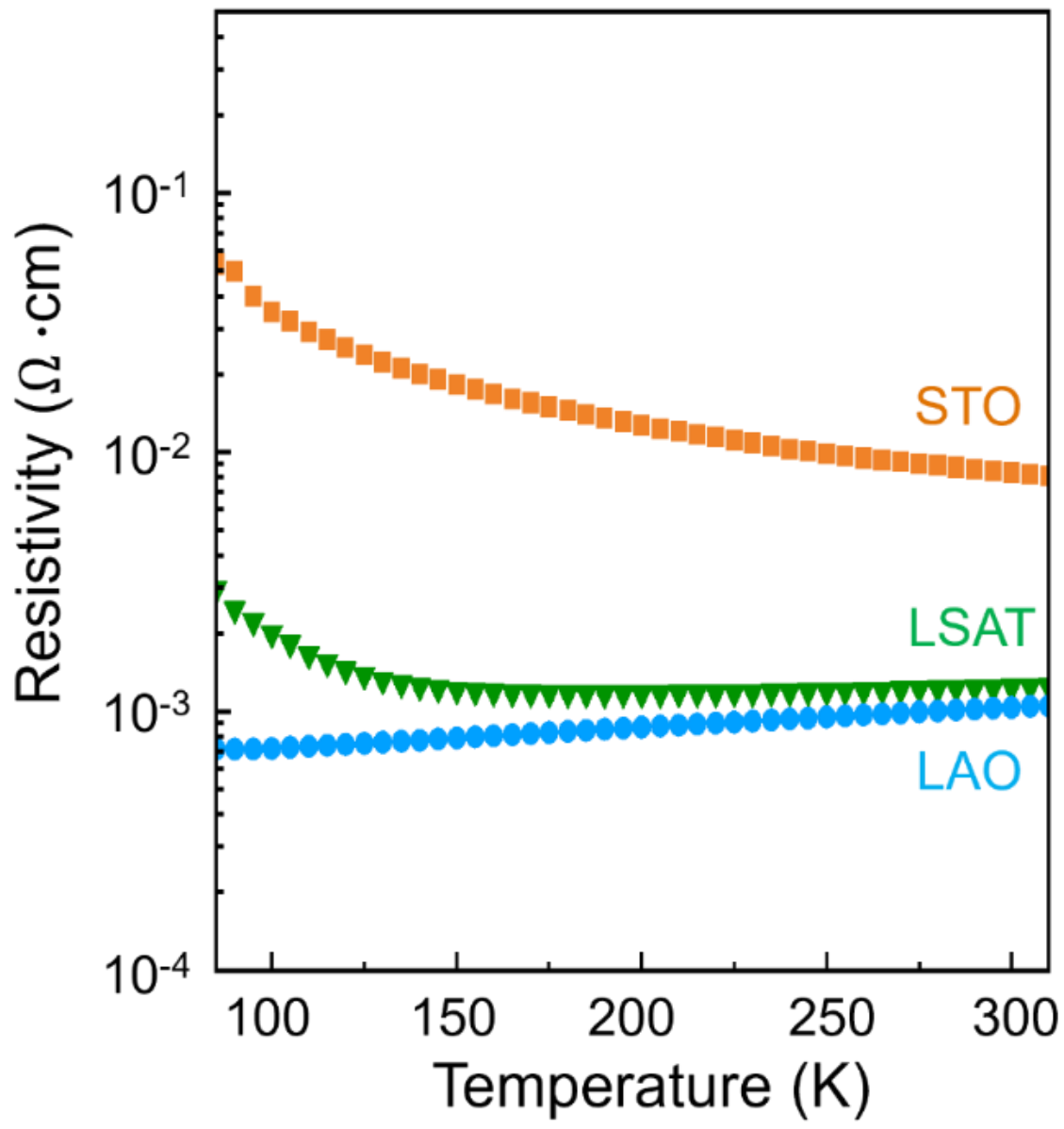

**Figure S4.** Resistivity–temperature curves of 13 nm thick (5A)$NiO_3$ films grown on LAO, LSAT, and STO.

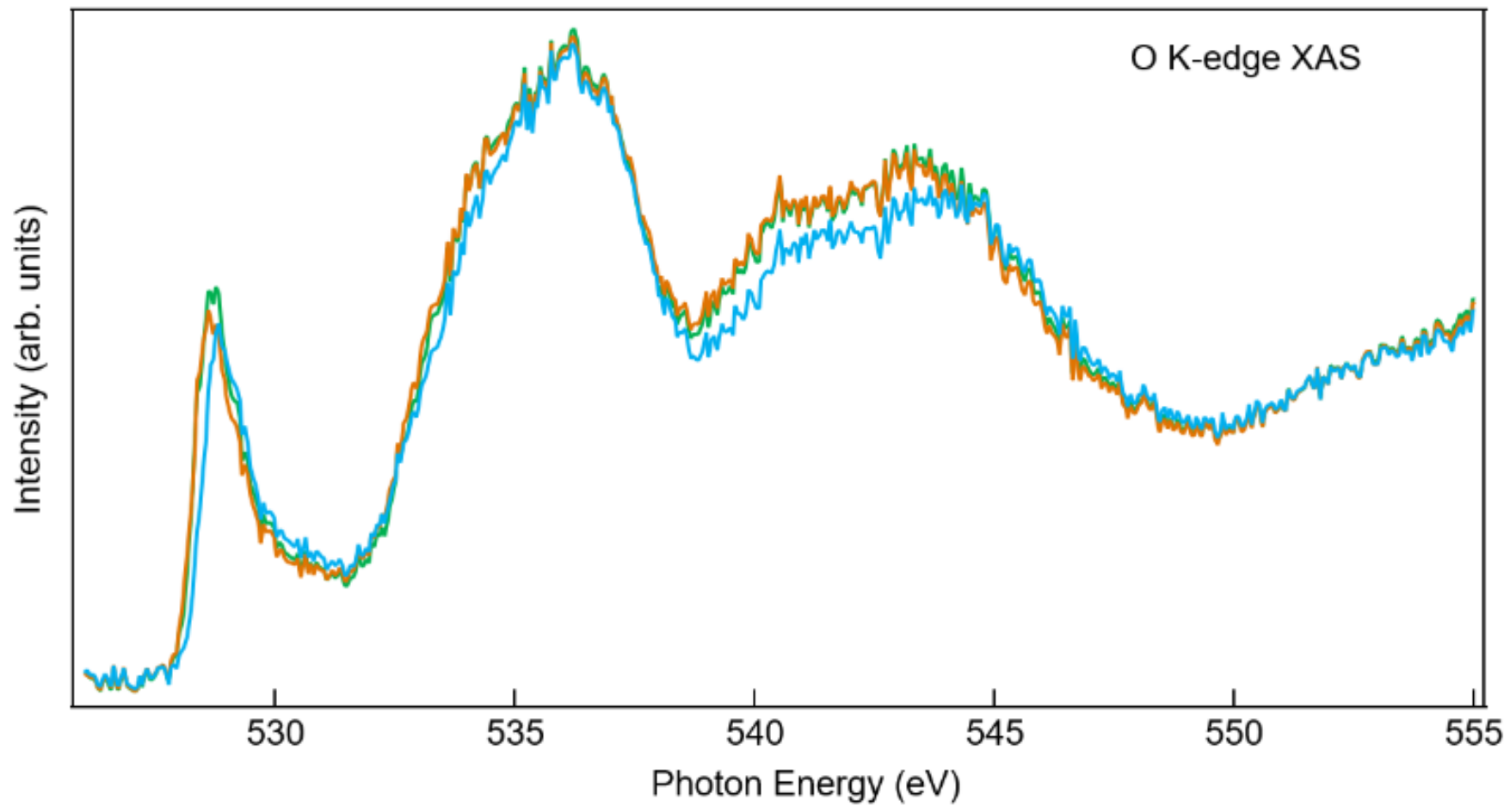

**Figure S5.** Full O K-edge XAS spectra of (5A)$NiO_3$ thin films.

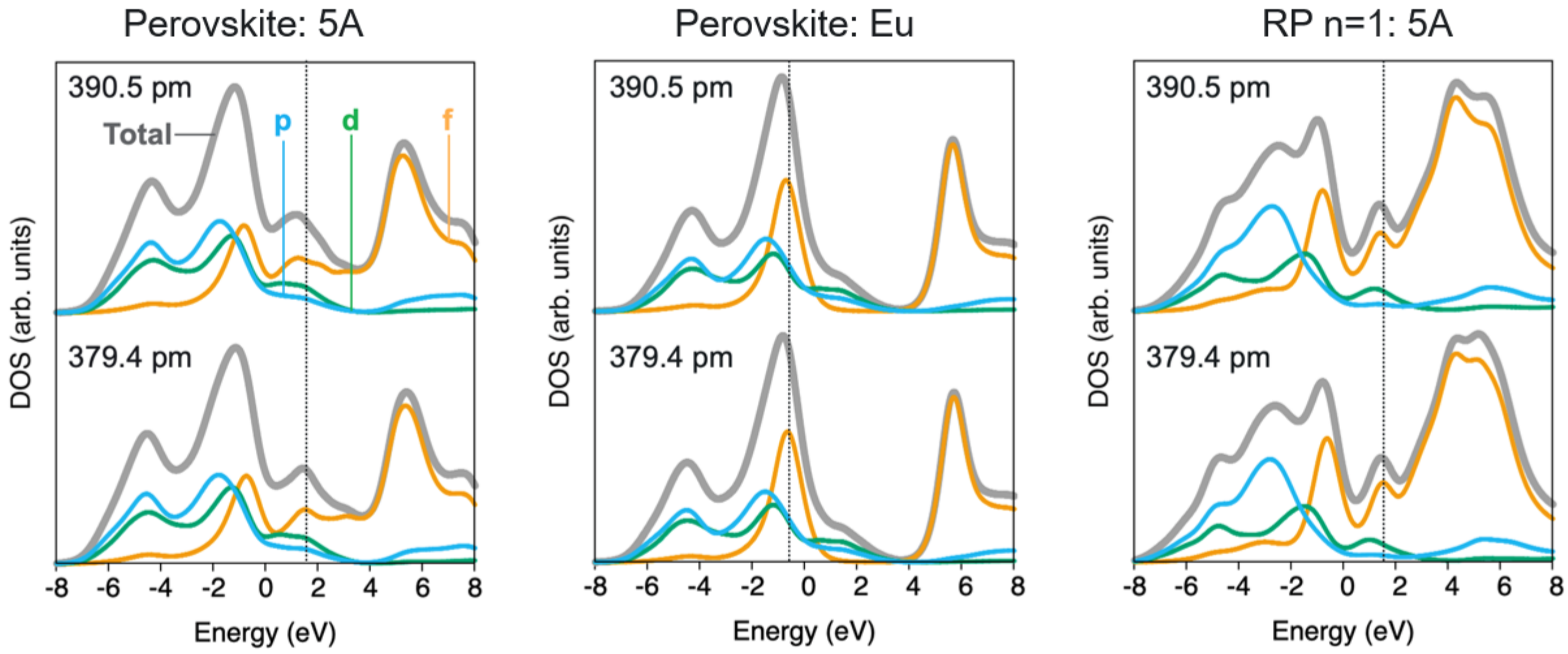

**Figure S6.** Strain-dependent calculated DOS for perovskite and RP-type nickelate structures. Calculated density of states for $(5A)NiO_3$ and $EuNiO_3$ perovskites and the $(5A)_2NiO_4$ RP n=1 phase under tensile (STO: 3.905 Å) and compressive (LAO: 3.794 Å) in-plane lattice parameters. The total DOS is shown in gray, with projected contributions from p, d, and f orbitals shown in blue, green, and orange, respectively. These contributions are dominated by O, Ni, and A-site rare-earth states, respectively; the s-orbital contribution is negligible in the plotted energy range. The Fermi energy is set to 0 eV. Vertical dashed lines are guides to the eye, highlighting the strain-dependent shift of selected unoccupied states. In the $(5A)NiO_3$ perovskite, tensile strain shifts the unoccupied f-derived states near the Fermi level to lower energy by approximately 0.5 eV, whereas this shift is weaker in the RP phase. The occupied f states and higher-energy unoccupied states show no obvious strain-dependent shift.

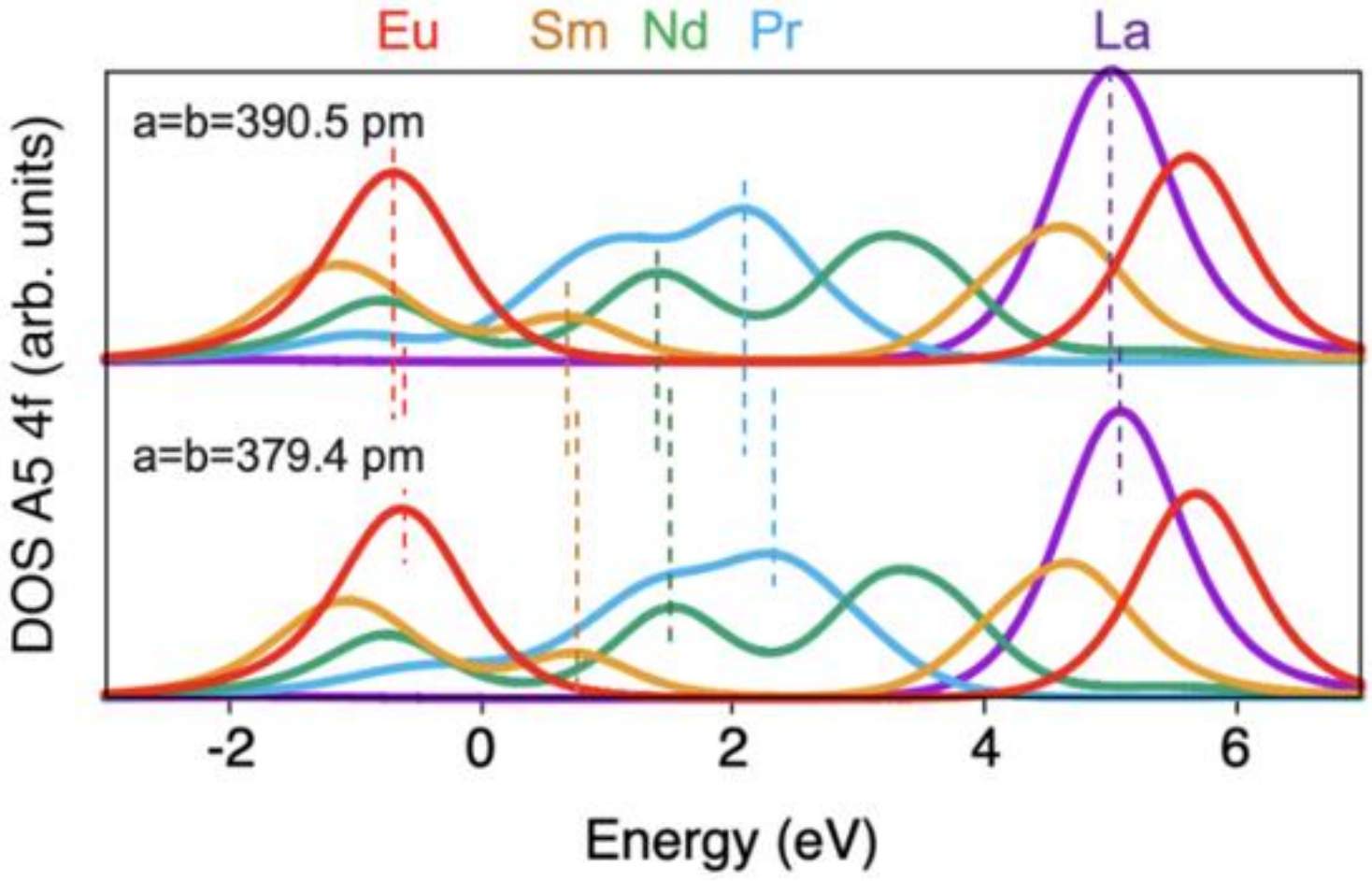

**Figure S7.** Strain dependence of the rare-earth 4f-projected DOS in perovskite (5A)$NiO_3$. Calculated density of states projected onto the 4f orbitals of the 5A elements: La (purple), Pr (light blue), Nd (green), Sm (orange), and Eu (red). The Fermi energy is set to 0 eV, and vertical dashed lines are guides to the eye. Tensile strain shifts the unoccupied 4f states to lower energies for all rare-earth elements, with the strongest shift observed for Pr.

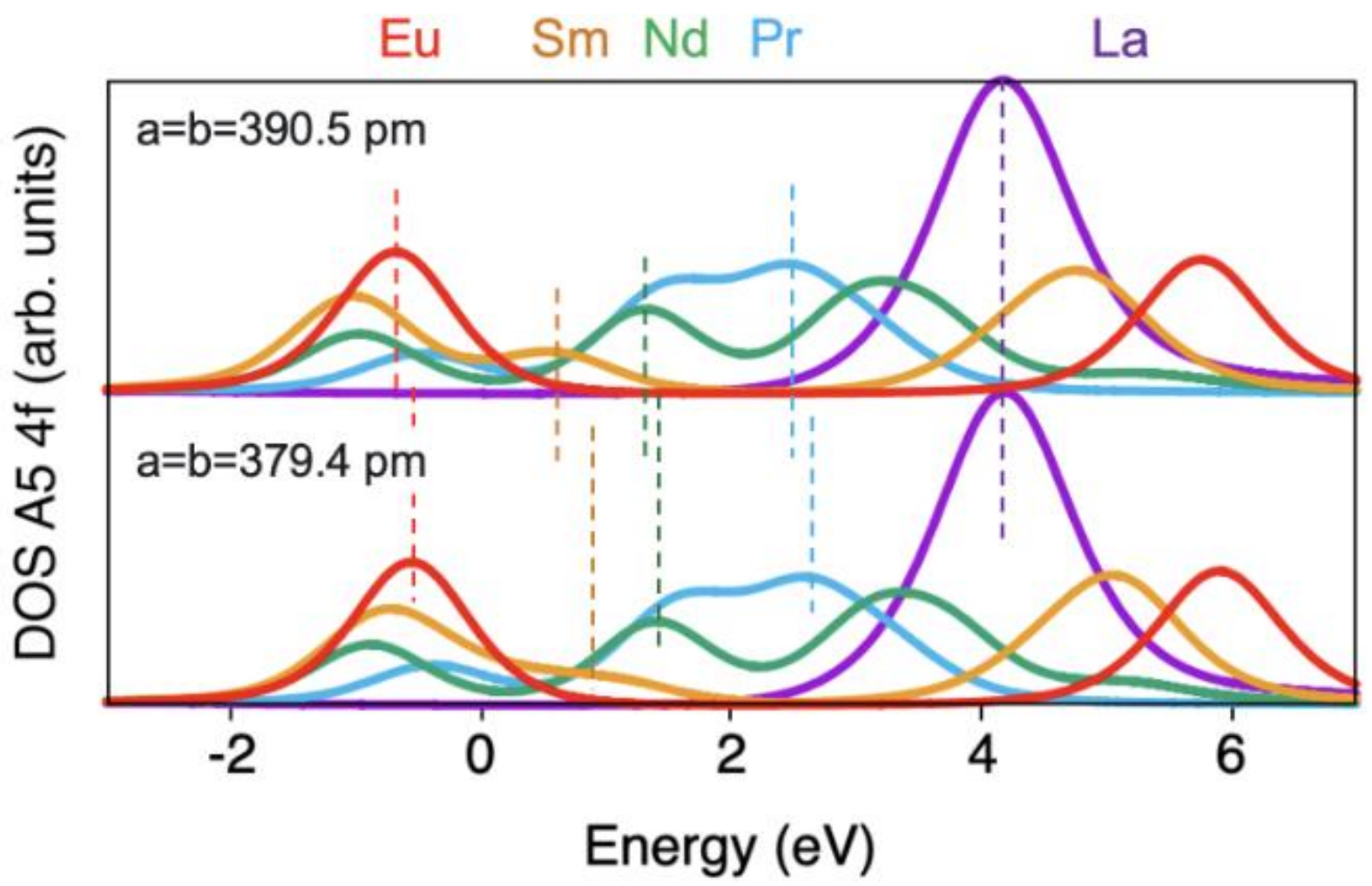

**Figure S8.** Strain dependence of the rare-earth 4f-projected DOS in RP-type $(5A)_2NiO_4$. Calculated density of states projected onto the 4f orbitals of the 5A elements in the RP-type $(5A)_2NiO_4$ structure: La (purple), Pr (light blue), Nd (green), Sm (orange), and Eu (red). The Fermi energy is set to 0 eV, and vertical dashed lines are guides to the eye. Tensile strain shifts the unoccupied 4f states to lower energies for all rare-earth elements, with the strongest shifts observed for Eu and Sm.

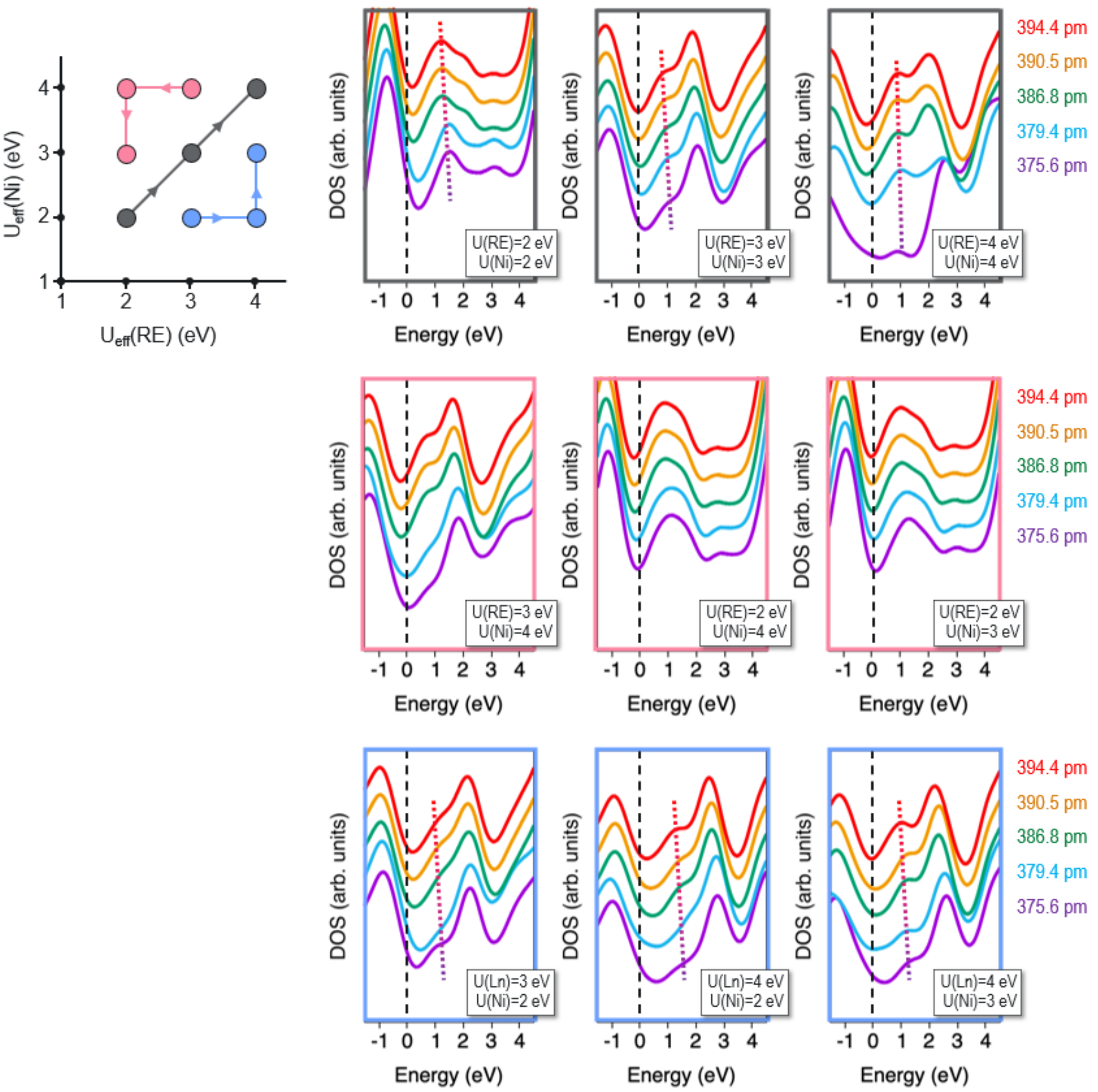

**Figure S9.** Top left: Selected values of the $U_{eff}$(RE) and $U_{eff}$(Ni) corrections used to examine the effect of the Hubbard U on the dependence of the f-projected DOS on the epitaxial strain. Gray, **pink**, and **blue** indicate three types of conditions: $U_{eff}(RE)=U_{eff}(Ni)$, $U_{eff}(RE)<U_{eff}(Ni)$, $U_{eff}(RE)>U_{eff}(Ni)$, respectively. Right: The calculated DOS projected on 4f states of all RE atoms. Comparing the DOS for the $U_{eff}(RE)= U_{eff}$ (Ni) and $U_{eff}$ (RE)> $U_{eff}$ (Ni) conditions shows that the strain-induced shift of the lowest-energy peak (indicated with dotted lines) is independent of U. However, as $U_{eff}$ (RE) increases, the DOS intensity near the Fermi level (indicated with vertical dashed lines) shows strong strain dependence. Both factors suggest that the RE 4f DOS shifts to lower energy with tensile strain. We note that the U values applied to the lanthanide 4f states are at the lower end of values commonly used for lanthanides (typically ~5–6.5 eV);[1] nevertheless, the effect of strain is already clearly resolved at U as low as 2 eV. We also note that the selected range of U(Ni) covers the range of typical values for multicomponent nickelate compounds.[2-4]

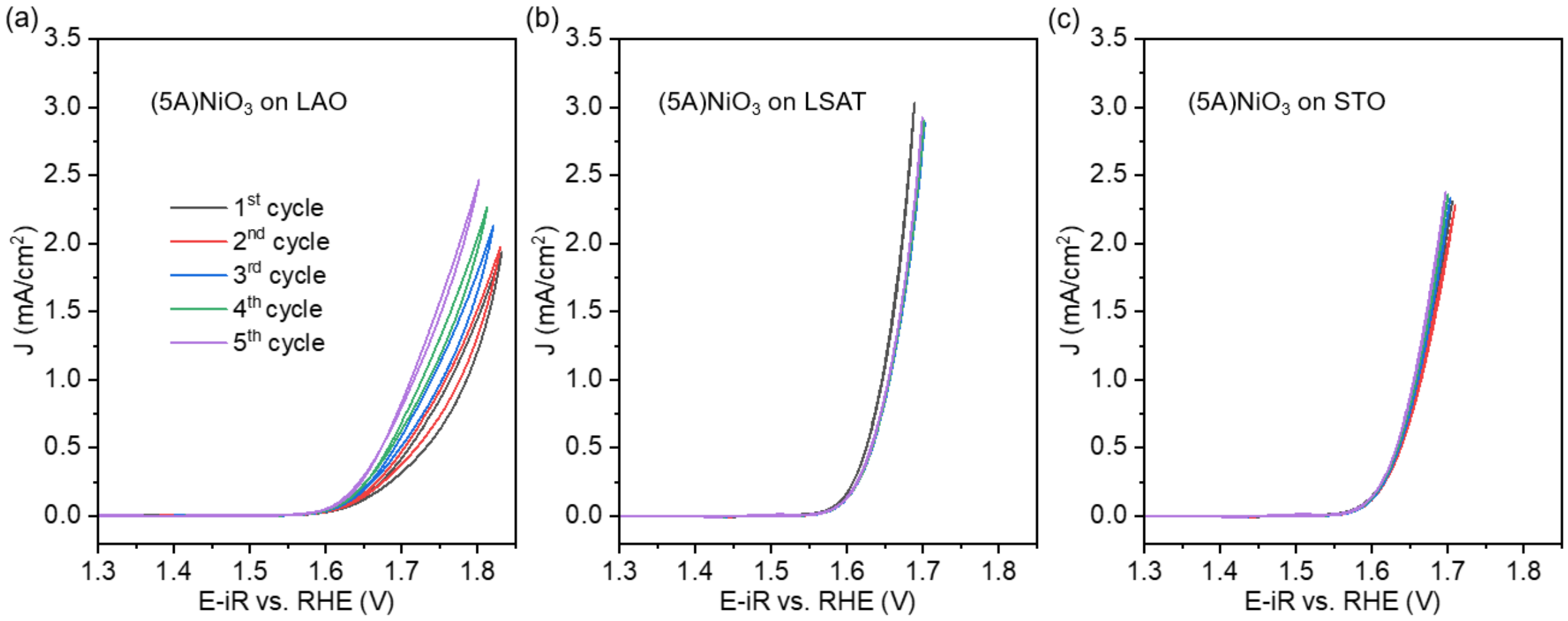

**Figure S10. CV cycling behavior of (5A)$NiO_3$ thin films during OER.** First to fifth iR-corrected CV cycles of (5A)$NiO_3$ films grown on (a) LAO, (b) LSAT, and (c) STO, measured in $O_2$-saturated 0.1 mol/L KOH. The LAO film shows gradual activation with cycling, whereas the tensile-strained LSAT and STO films exhibit relatively stable OER polarization curves over the first five cycles.

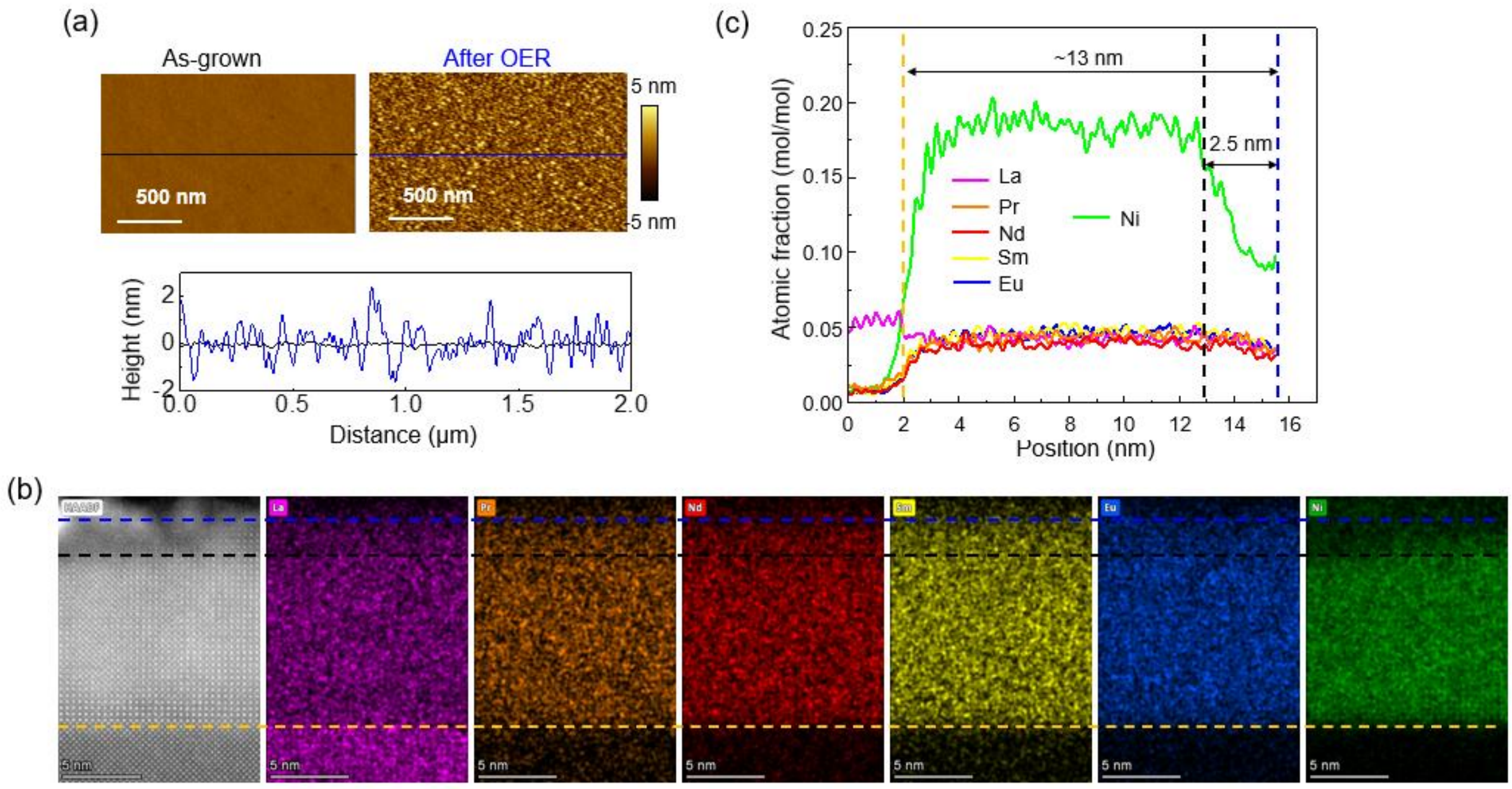

**Figure S11.** Post-OER morphological, structural, and compositional characterization of the (5A)$NiO_3$ thin film on LSAT. (a) AFM images of the as-grown and post-OER films, together with a representative surface height profile comparison. (b) Cross-sectional HAADF-STEM image and corresponding EDS elemental maps of La, Pr, Nd, Sm, Eu, and Ni after OER operation. (c) Quantitative EDS atomic-fraction profiles across the film thickness, showing relatively uniform distributions of the five rare-earth cations in the film interior, with approximately equimolar A-site proportions. A near-surface region of approximately 2.5 nm exhibits a pronounced decrease in the measured Ni atomic fraction and a smaller reduction in the rare-earth cations. The yellow, black, and blue dashed lines indicate the film/substrate interface, the approximate boundary of the near-surface modified region, and the film surface estimated by the thickness of 13 nm, respectively.